\documentclass[10pt,final,journal]{IEEEtran}
\usepackage{amsmath,graphicx}
\usepackage[utf8]{inputenc}
\usepackage{tabulary}
\usepackage{adjustbox}
\usepackage{graphicx}
\usepackage{multirow,array}
\usepackage{hhline}
\usepackage{amsmath}
\usepackage{amssymb}
\usepackage{rotating}
\usepackage{xcolor}
\usepackage{balance}
\usepackage{indentfirst}
\usepackage{makecell}
\usepackage{blindtext}
\usepackage{mathtools}

\usepackage[caption=false,font=footnotesize,labelfont=sf,textfont=sf]{subfig}

\usepackage{comment}
\begin{document}

\title{Harmonic and non-Harmonic Based Noisy Reverberant Speech Enhancement in Time Domain}

\author{G. Zucatelli, ~\IEEEmembership{Student Member,~IEEE}
        and R. Coelho,~\IEEEmembership{Senior Member,~IEEE}

\thanks{R. Coelho is partially supported by the National Council for Scientific and Technological Development (CNPq) 308155/2019 and Fundação de Amparo à Pesquisa do Estado do Rio de Janeiro (FAPERJ) 203075/2016 research grants. This work is also supported by the Coordenação de Aperfeiçoamento de Pessoal de Nível Superior - Brasil (CAPES) - Grant Code 001. 
The authors are with the Laboratory of Acoustic Signal Processing, Military Institute of Engineering (IME), Rio de Janeiro, RJ 22290-270, Brazil (e-mail: coelho@ime.eb.br).}}

\markboth{ }%
{Shell \MakeLowercase{\textit{et al.}}: Bare Demo of IEEEtran.cls for IEEE Journals}

\maketitle

\begin{abstract}
This paper introduces the single step time domain method named HnH-NRSE, whihc is designed for simultaneous speech intelligibility and quality improvement under noisy-reverberant conditions.
In this solution, harmonic and non-harmonic elements of speech are separated by applying zero-crossing and energy criteria. 
An objective evaluation of the its non-stationarity degree is further used for an adaptive gain to treat masking components.
No prior knowledge of speech statistics or room information is required for this technique.
Additionally, two combined solutions, IRMO and IRMN, are proposed as composite methods for improvement on noisy-reverberant speech signals.
The proposed and baseline methods are evaluated considering two intelligibility and three quality measures, applied for the objective prediction.
The results show that the proposed scheme leads to a higher intelligibility and quality improvement when compared to competing methods in most scenarios.
Additionally, a perceptual intelligibility listening test is performed, which corroborates with these results.  
Furthermore, the proposed HnH-NRSE solution attains SRMR quality measure with similar results when compared to the composed  IRMO and  IRMN techniques.
\end{abstract}


\IEEEpeerreviewmaketitle


\section{Introduction} 
\label{Intro}
\IEEEPARstart{S}{peech} communication in urban environments is an essential challenge in daily life.
In these scenarios, the speech signal is generally submitted to the presence of acoustic noises and reverberation. 
This situation may lead to masking effects on speech signals and can also alter the human auditory perception \cite{Bolt_1949}\cite{Nabelek_74}\cite{Tang_2018}\cite{Madmoni_2021}.
Consequently, noisy-reverberant interferences reduce the quality and the intelligibility of speech signals by inducing changes on relevant harmonic structures of speech \cite{lu2009contribution}\cite{brown2010fundamental}. 
{\color{black} Studies have demonstrated that intelligibility and quality are related \cite{Dong_2020}, which motivates combined solutions that simultaneously improve these characteristics under adverse conditions.}


{\color{black} Methods have been designed to exclusively improve speech intelligibility under noisy and reverberant conditions \cite{Odya_2021}\cite{Taal2014speech}\cite{Hodo_10}\cite{Petkov_16}\cite{Zuca_20}.}
Spectral solutions usually adopt energy equalization to deal with these acoustic distortions.
The goal is to change the speech spectral energy distribution in other to have a more intelligible signal propagated inside a room.
This is the case of the ACO (Adaptive Compressive Onset-Enhancement) \cite{Bederna_2020}, which alters the energy present in different frequency bands considering prior knowledge on the acoustic noise and the Room Impulse Response (RIR).
A similar approach based on the PDMSE (Perceptual Distortion Measure-based Speech Enhancement) \cite{Taal2014speech}  and FIR (Fast Inverse Filtering) techniques \cite{tokuno1997inverse} was proposed in \cite{Dong_2018} with interesting objective and subjective intelligibility results.
In \cite{Niermann_2021} two solutions are described to cope with acoustic noise in order to improve speech intelligibility by spectral equalization strategies.
Temporal adaptive methods focus on preserving speech transient regions that are  considered more relevant for speech intelligibility due to greater phonetic information \cite{Strange_1983}.
The SSS (Steady-State Suppression) \cite{Hodo_10} was proposed considering the suppression of steady-state components that usually indicates frames with greater energy.
By amplitude attenuation, there is a reduction on the masking distortion that affects transient regions, therefore increasing the intelligibility. 
Another temporal approach is the AGC (Adaptive Gain Control) \cite{Petkov_16}, that adjusts the energy of each frame in order to reduce masking effects on relevant speech regions by an optimization criteria.
In the ARA\textsubscript{NSD} (Adaptive Reverberation Absorption using Non-Stationarity Detection) method \cite{Zuca_20} the non-stationarity of speech regions is accessed by the INS (Index of Non-Stationarity) \cite{Flandrin_10}.
This solution is able to restore the natural non-stationary behavior of speech signals.
Furthermore, it also attained an interesting speech quality gain \cite{Zuca_20}. 
This method defines the RGs (Reverberation Groups) regions in order to observe the non-stationary behavior of speech signals and absorbs masking components based on the non-stationarity variation.

In the literature, speech enhancement approaches are designed to mainly improve the speech quality in adverse conditions \cite{Cohen_01}\cite{Zao_14}\cite{zao_15}\cite{Tavares_16}.
These solutions identify the distortion components based on temporal and spectral estimators \cite{Norholm_16} or using decomposition techniques such as the EMD (Empirical Mode Decomposition) \cite{Huang_98}.
The OMLSA (Optimally-Modified Log-Spectral Amplitude) \cite{Cohen_01} adopts an acoustic noise estimator to access the spectral noise power and reconstruct the speech signal based on the minimization of the log-spectral mean square error. 
Another speech enhancement, the EMDH (Empirical Mode Decomposition with Hurst exponent) \cite{Zao_14}\cite{zao_15} is a temporal method that employs the Hurst exponent \cite{Hurst_51} to identify and remove Intrinsic Mode Functions (IMFs) most corrupted by noise from the EMD decomposition.
The NNESE (Nonstationary Noise Estimation for Speech Enhancement) strategy \cite{Tavares_16} is a temporal solution that employ the robust DATE estimator \cite{Pastor_12} on a short-time basis, being able to account for the non-stationarity of acoustic signals.

In order to simultaneously improve speech intelligibility and quality, the TSDL (Two-Stage with Deep Learning) \cite{Wang_2019} approach is proposed based on two DNN (Deep Neural Network) stages.
Each stage is trained in to separately and consecutively mitigate noise and reverberation for concurrent improvement.
The combined approach presented in \cite{IBM2007} is a composition of the acoustic mask IBM (Ideal Binary Mask) \cite{IBM2007} and the OMLSA speech enhancement technique \cite{Cohen_01} to provide intelligibility and quality improvement in noisy environments.

In this paper, the method HnH-NRSE (Harmonic and non-Harmonic based Noisy-Reverberant Speech Enhancement) is proposed as a single approach to attain simultaneous intelligibility and quality improvement of noisy-reverberant speech signals.
The solution takes into account the non-stationarity variations of speech for the adaptive gain approach and consider an harmonic and non-harmonic decomposition, since the harmonic components are naturally related to the intelligibility \cite{brown2010fundamental}\cite{hong2010detection}. 
Additionally, two combined solutions IRMO (Ideal Reverberant Mask with OMLSA) and IRMN (Ideal Reverberant Mask with NNESE) are proposed for the noisy-reverberant scenario.
These methods adopt the Ideal Reverberant Mask (IRM) \cite{Loizou_11} in composition with speech enhancement techniques OMLSA and NNESE, and is also designed for both intelligibility and quality improvement.


Extensive experiments are conducted to objectively evaluate the proposed solutions for speech intelligibility and speech quality gain.
The noisy-reverberant scenario is composed of two real reverberant rooms from the AIR \cite{Air_09} and LASP\_RIR\footnote{Available at www.lasp.ime.eb.br} databases and two background non-stationary acoustic noises extracted from RSG-10 \cite{Steen_88} and DEMAND \cite{Thie_2013} databases.
The intelligibility assessment in performed considering the objective measures STOI \cite{Taal_11} and ASII\textsubscript{ST} \cite{Hen_15}.
The SRMR \cite{Falk_14}, PESQ \cite{Pesq_01} and f2-model PEAQ \cite{PEAQ_1999} \cite{PEAQ_2fmodel} measures are adopted for quality objective evaluation.
A subjective intelligibility listening test is also performed and results show that the proposed HnH-NRSE method outperforms competing techniques in terms of speech intelligibility. 
Moreover, the HnH-NRSE is able to achieve similar SRMR values when compared to the proposed combined approaches.  
  
The main contributions of this work are:
\begin{itemize}
 \item introduction of the HnH-NRSE single method for simultaneous intelligibility and quality improvement of noisy-reverberant speech signals in time domain.
 \item design of composite solutions IRMN for noisy-reverberant speech signals improvement.
 \item definition of composite solution IRMO for speech improvement under noisy-reverberant environments.
\end{itemize}

The remainder of this paper is organized as follows. 
In Section II, the noisy-reverberant effect is defined and analyzed considering the speech harmonic structure and non-stationarity. 
The single stage HnH-NRSE solution and proposed combined methods IRMO and IRMN are introduced in Section III.
The objective measures used to evaluate the techniques on speech quality and intelligibility are briefly described in Section IV. 
The experiments are exposed in Section V.
Then, the results for improvements on speech intelligibility and quality are presented and discussed for all methods.
Finally, Section VI concludes this work.







\section{Noisy-Reverberant Acoustic Effects} 
The reverberation effect is usually defined as a linear filtering process such that, given a RIR $h(t)$, the reverberated signal can be obtained by convolution. 
In real environments, acoustic noises are also a common distortion, which means that the resultant noisy-reverberant speech signal $r(t)$ can be obtained by 
\begin{equation}
r(t) = s(t)*h(t) + w(t),
\end{equation}
where $s(t)$ is the clean speech signal and $w(t)$ is the background noise.




\begin{figure}[t]
    \centering
    \vspace{.3cm}
    \includegraphics[width=4.4cm]{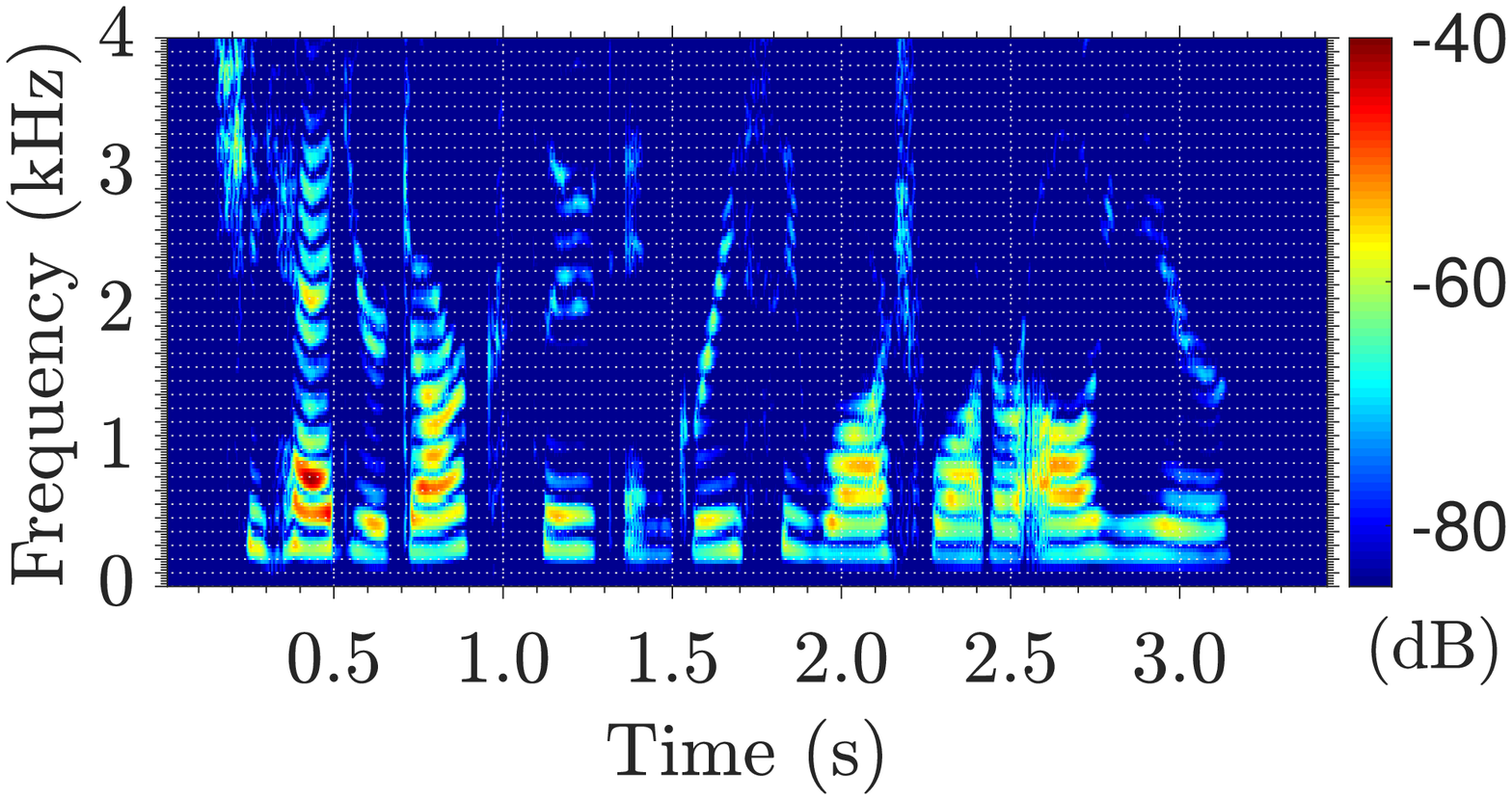}
    \includegraphics[width=4.2cm]{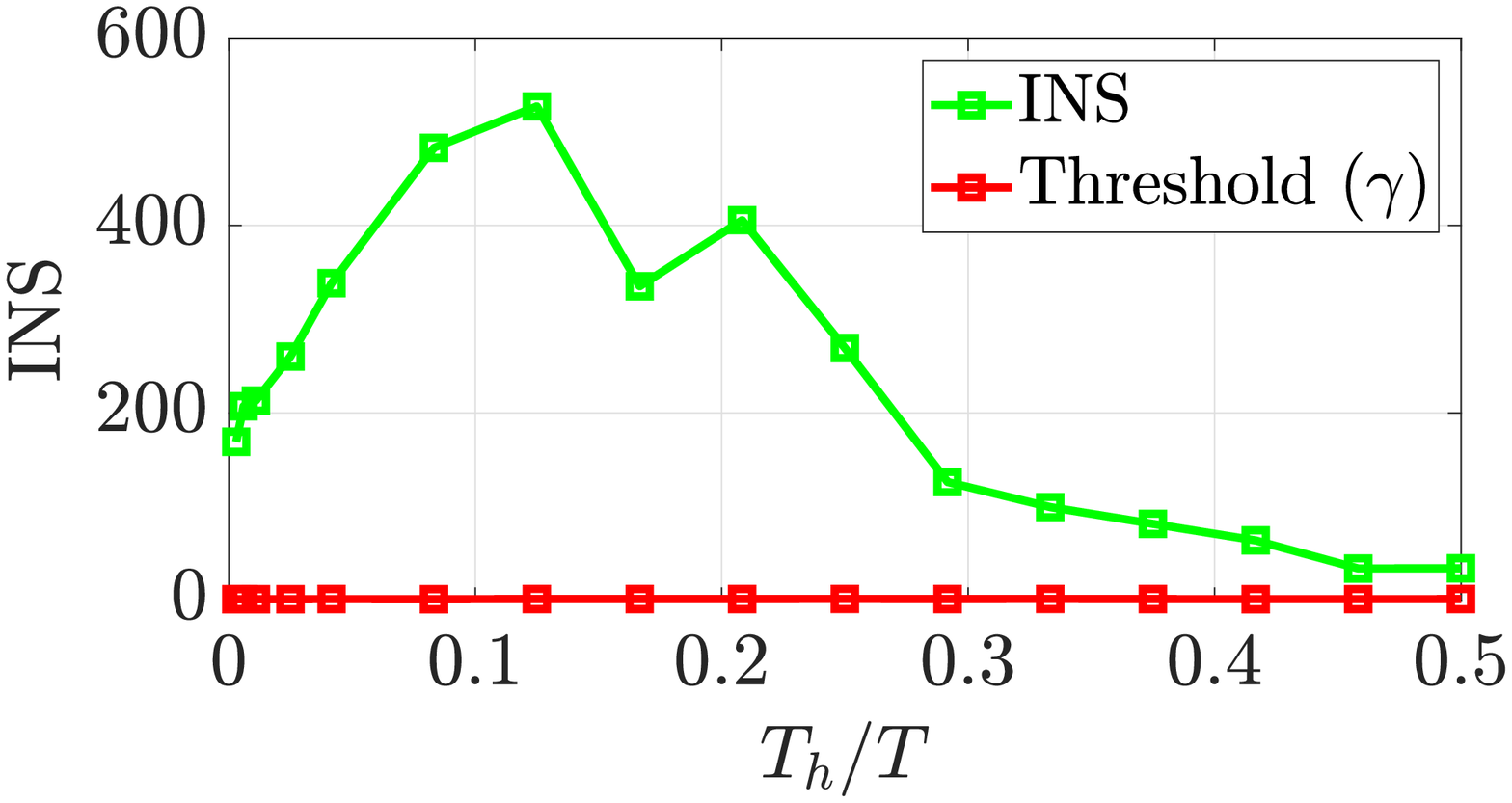} \\
    {\small  \vspace{-0.15cm} (a)} \\ \vspace{0.1cm} %
    
    \includegraphics[width=4.4cm]{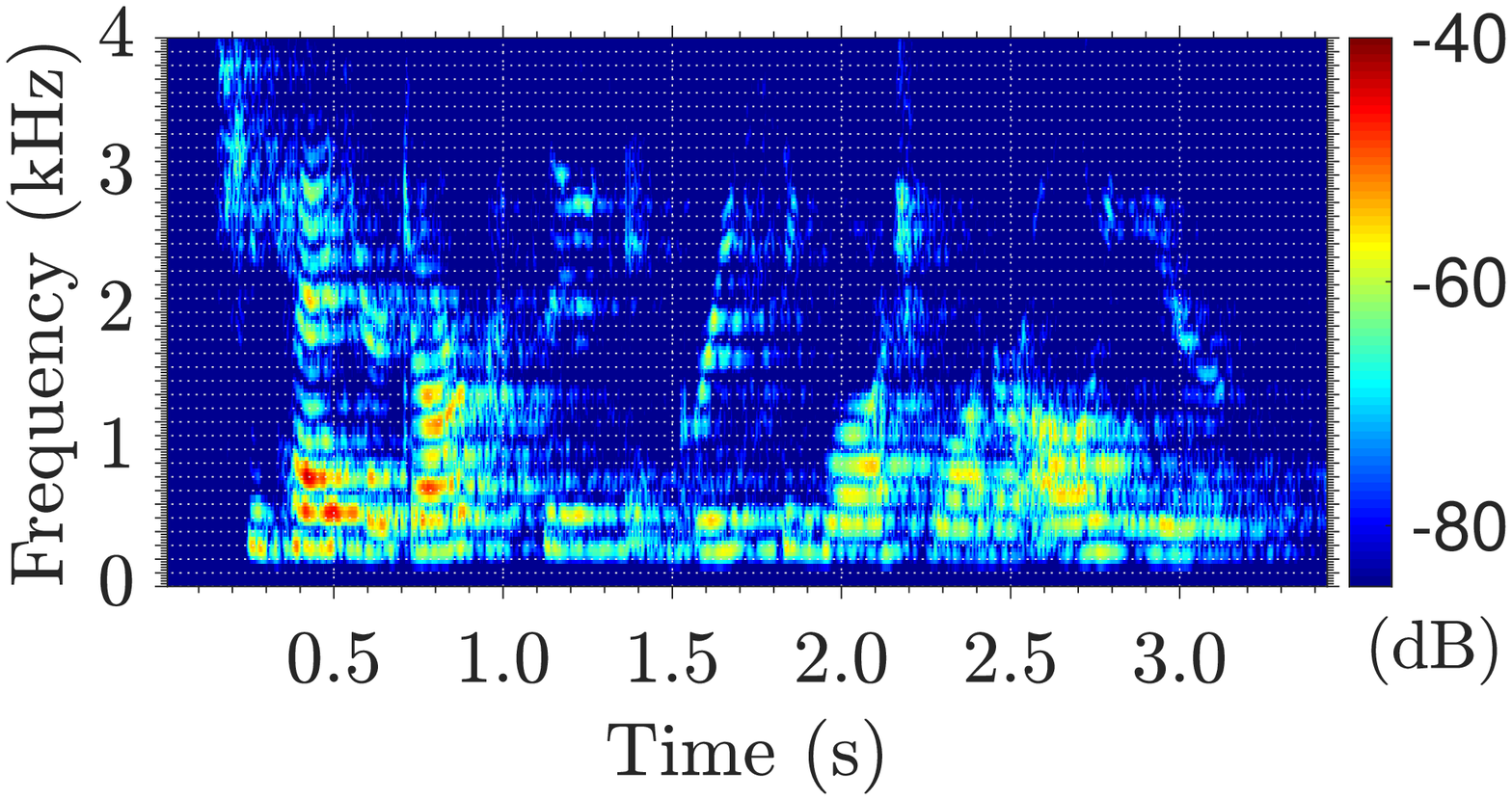}
    \includegraphics[width=4.2cm]{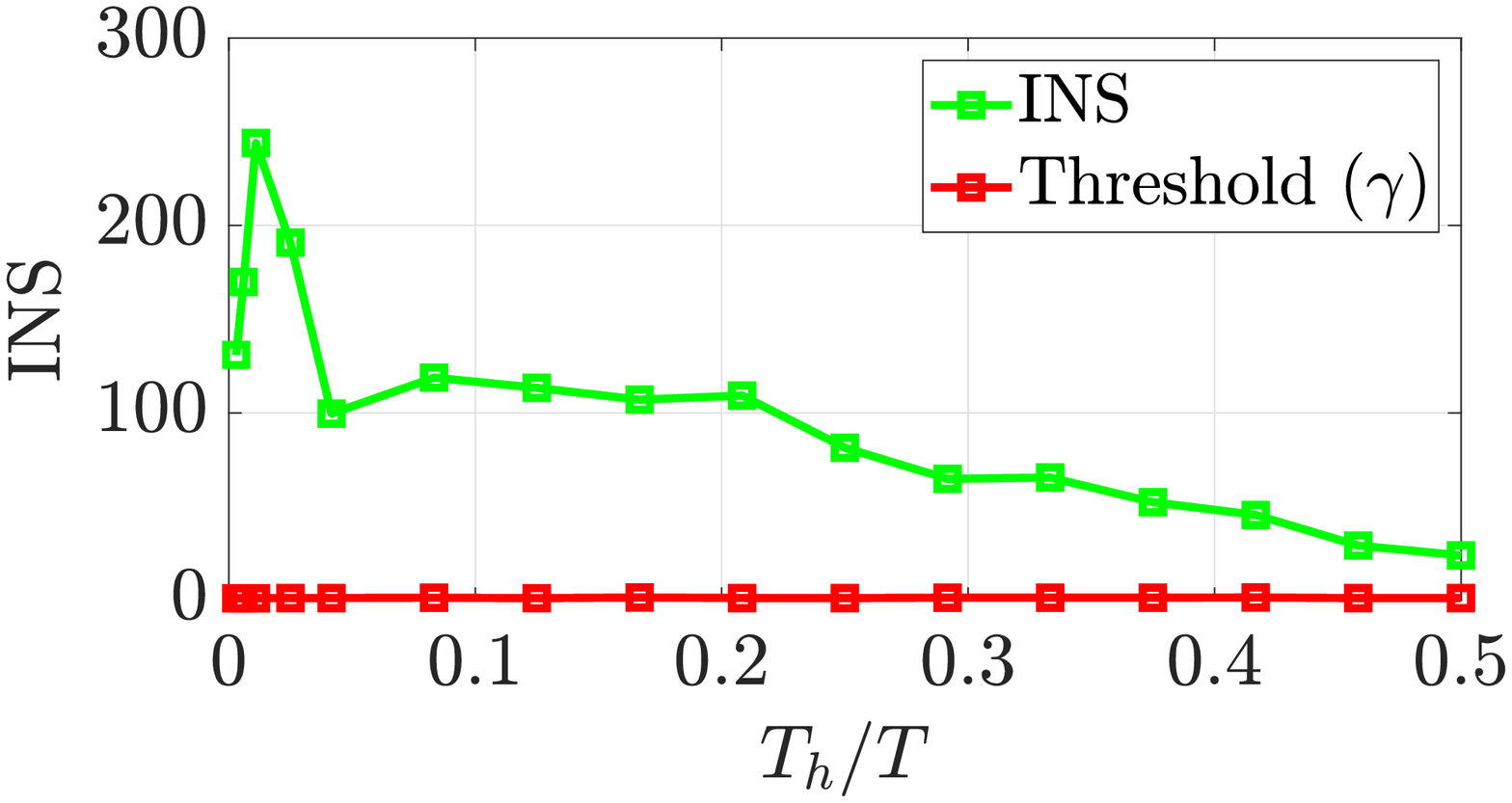} \\
    {\small  \vspace{-0.15cm} (b)} \\ \vspace{0.1cm} %
    
    \includegraphics[width=4.4cm]{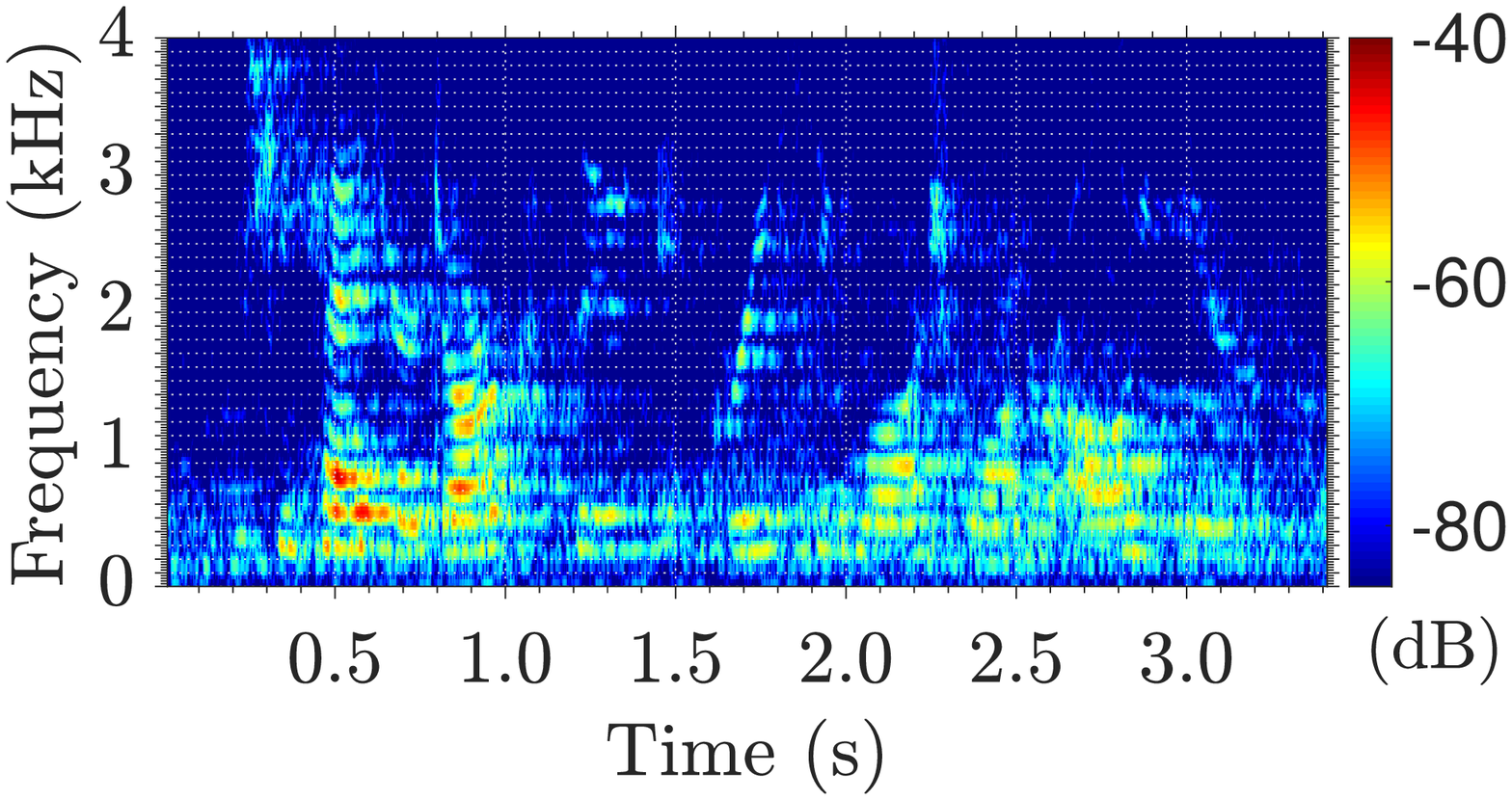}
    \includegraphics[width=4.2cm]{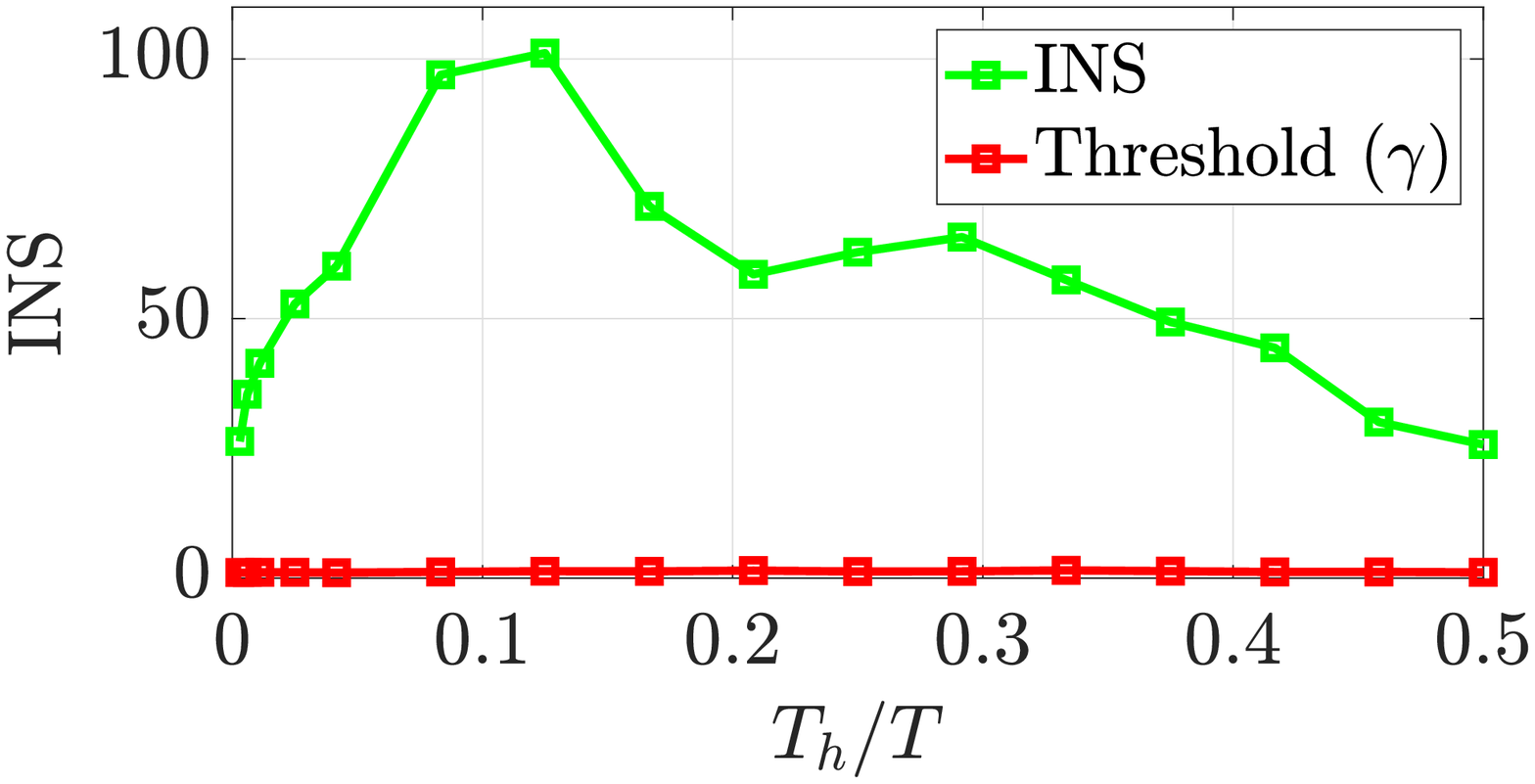} \\
    {\small  \vspace{-0.15cm} (c)} \\ \vspace{0.1cm} %
    \caption{Spectrogram and INS of clean (a), reverberant (b) and noisy-reverberant (c) speech signals for LASP2 room and Babble noise at $10$ dB.}
    \label{Fig::Specs}
    \vspace{.2cm}
\end{figure}

Fig. \ref{Fig::Specs} depicts the spectrogram and INS \cite{Flandrin_10}  of the clean, reverberant and noisy-reverberant speech signal considering the LASP2 room\footnotemark[1] and the Babble noise at $10$ dB.
It can be seen that the distinction between silence and voice active regions present on the direct signal is masked by the noisy and reverberant effects.
Moreover, the harmonic structure of the speech signal is altered, changing observable formants, specially for lower frequencies.  
This illustrate how the noisy-reverberant effect alter the harmonic structure of the speech signals, change human auditory perception  and reduce acoustic intelligibility and quality \cite{Bolt_1949}\cite{Nabelek_74}\cite{brown2010fundamental}.
Moreover, the corresponding INS values show how the natural non-stationary behavior of speech signals is altered in these conditions.

\begin{figure*}[t!]
\centering
\includegraphics[width=15cm]{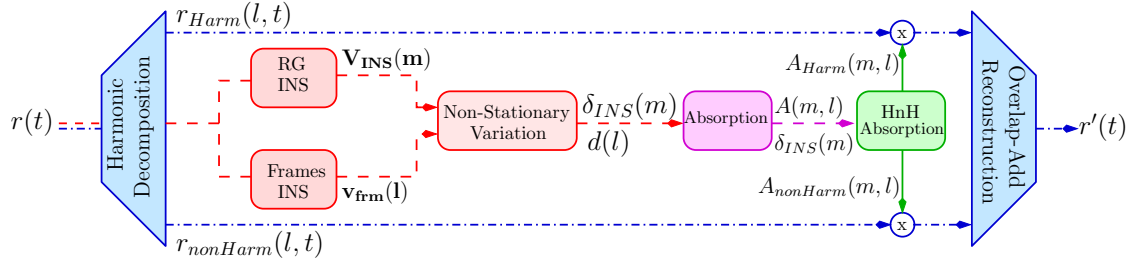}
\caption{Block diagram for the proposed HnH-NRSE method. The harmonic structure of the speech signal is adopted as a criteria for signal decomposition and adaptive absorption of noisy-reverberant masking components.}
\label{Fig::Scheme_HnH}
\vspace{-.3cm}
\end{figure*}

This objective measure compares the target signal with stationarity references called surrogates for different time scales $T_h/T$, where $T_h$ is the short-time spectral analysis length and $T$ is the total signal duration.
Surrogate signals have the same absolute spectrogram values of a target signal and phase identically distributed in $[-\pi,\pi]$ \cite{Flandrin_10}. 
For each length $T_h$, the INS is defined using the ratio between the variance of spectral distances from local to global multitaper spectrograms for the target signal ($\Theta_1$) and corresponding surrogates ($\Theta_0(k)$) as
\begin{equation}
\text{INS} \coloneqq \sqrt{\frac{\Theta_1}{\langle \Theta_0 (k) \rangle}_k}.
\end{equation}
A threshold $\gamma$ is defined to keep the stationarity assumption considering a $95\%$ confidence degree as
\vspace{-.05cm}
\begin{equation}
 INS \quad \left\{\begin{matrix}
 \hspace{-.75cm} \leq \gamma, \quad \text{signal is stationary} \\
 >  \gamma, \quad \text{signal is non-stationary}. 
\end{matrix}\right.
\end{equation}
The harmonic components of the speech signal are altered by both masking effects  that reduce the natural non-stationary behavior of speech signals from a maximum value of $600$ down to $300$ and $100$ for the reverberant and noisy-reverberant scenarios, respectively.



\section{Proposed Methods} 
In this Section, the HnH-NRSE method is presented, which aims to simultaneously improve intelligibility and quality of noisy-reverberant speech signals.
Moreover, the two composed solutions IRMN and IRMO establish the intrusive reference as a composition of the Ideal Reverberant Mask (IRM) \cite{Loizou_11} and speech enhancement methods NNESE \cite{Tavares_16} and  OMLSA \cite{Cohen_01}.

\subsection{Proposed HnH-NRSE: Harmonic and non-Harmonic based Noisy Reverberant Speech Enhancement}
The HnH-NRSE aims to apply the adaptive absorption based on variations of speech non-stationarity \cite{Zuca_20} taking into account the harmonic elements of speech signal.
A block diagram is illustrated in Fig. \ref{Fig::Scheme_HnH}.
{\color{black} The solution can be described in three main steps: harmonic and non-harmonic decomposition, adaptive masking absorption and signal reconstruction.}

\subsubsection{Harmonic and non-Harmonic Separation}
Given a noisy-reverberant signal $r(t)$, consider its $l$-th frame representation $r(l,t)$ and the maximum energy $EN_{\text{max}}$ observed in all frames.
The harmonic and non-harmonic separation is performed based on the energy and zero-crossing thresholds \cite{Rabiner_1978}, $EN_{th}$ and $ZC_{th}$, as
\begin{equation}
 r_{\text{Harm}}(l,t) = \begin{cases}
                        r(l,t), \quad ZC(l)<ZC_{th} \text{ and } \\ 
                        \quad \quad \quad \quad EN(l)-EN_{\text{max}}>EN_{th} \\
                        [0 \dots 0], \quad \quad \text{otherwise,}
                    \end{cases}
\end{equation}
and
\begin{equation}
r_{\text{non-Harm}}(l,t) = \begin{cases}
                        [0 \dots 0], \quad ZC(l)<ZC_{th} \text{ and } \\
                        \quad \quad \quad \quad EN(l)-EN_{\text{max}}>EN_{th} \\
                        r(l,t), \quad \quad \text{otherwise,}
                    \end{cases}
\hspace{-.2cm}
\end{equation}
where $EN(l)$ and $ZC(l)$ are the energy and zero-crossing values for a given frame, respectively.
For low SNR scenarios, when $EN_{\text{max}}-EN_{\text{min}}$ is greater than $EN_{th}$, the energy threshold is updated to $EN_{th}=-0,55*(EN_{\text{max}}-EN_{\text{min}})$.
This is important to achieve a better discrimination between harmonic and non-harmonic frames.

\subsubsection{Adaptive Masking Absorption}
For the adaptive absorption of masking components, it is adopted a short-time absorption procedure based on variations of the non-stationarity of Reverberation Groups (RG) \cite{Zuca_20} assessed by the INS \cite{Flandrin_10}.

The non-stationarity variation for consecutive RGs is defined by the normalized distance $\delta_{INS}(m) = ||{\mathbf{ v_{INS}}}(m)-{\mathbf{v_{INS}}}(m-1)||/(||{\mathbf{v_{INS}}}(m)||+||{\mathbf{v_{INS}}}(m-1)||)$, where $m$ stands for the $m$-th RG of a signal and $\mathbf{v_{INS}}$ its INS vector.
A similar approach is also performed on a frame-by-frame bases as $d(l) \in [0,1]$ to define the frame absorption $A(m,l)$ which is dependent on a non-stationarity threshold $\theta_{INS}$ as
\begin{equation}
\small 
A(m,l) = \begin{cases}
        F(l). \frac{L(m) - S}{1 + \exp(-k.(d(l)-d_0))} + S, \quad \delta_{INS} \leq \theta_{INS}; \\
        \frac{L'}{1 + \exp( -k'.(d(l) - d'_0 )}, \quad \hspace{1.3cm} \delta_{INS} > \theta_{INS},
        \end{cases}
\hspace{-.7cm}
\end{equation}
where $d_0$ and $d'_0$ are the inflection points with growth rates of $k$ and $k'$, $S$ is a minimum shift in order to avoid total absorption.
$L'$ and $L(m)= p\delta_{INS} + (1-p)L(m-1)$ define the maximum absorption values, where $p$ assigns the importance of the present RG and $F(l)$ is a factor defined in \cite{Zuca_20} such that $A(m,l) \approx L(m)$ only for $d(l) \approx 1$. 

Considering the harmonic structure of speech signals, the harmonic absorption approach $A_{\text{HnH}}(m,l)$ is defined as 
\begin{equation}
\small
 A_{\text{HnH}}(m,l) = \begin{cases}
            (1+f_{\text{HnH}} . (\delta_{INS})^{c_{\text{HnH}}}  ) A(m,l), \quad \text{harmonic}  \\
            (1-f'_{\text{HnH}} . (\delta_{INS})^{c'_{\text{HnH}}} ) A(m,l), \quad \text{non-harmonic},
        \end{cases}
\end{equation}
where $f_{\text{HnH}}$, $f'_{\text{HnH}}$, $c_{\text{HnH}}$ and $c'_{\text{HnH}}$ are free parameters adopted to incorporate the harmonic and non-harmonic information of each frame based on the non-stationary variation $\delta_{INS}$.
This is performed in order to enhance harmonic frames that are relevant for speech intelligibility and reduce noisy-reverberant masking components.
Therefore, the processed frame $r'(l,n)$ can be obtained by $r'(l) = Abs_{\text{HnH}}(m,l)*r_\text{Harm}(l,n) + Abs_{\text{HnH}}(m,l)*r_\text{não-Harm}(l,n)$.

\subsubsection{Speech Signal Reconstruction}
The last step of the HnH-NRSE method is the reconstruction of the speech signal as a processed version $r'(t)$.
To this end, resulting frames $r'(l,n)$ are used on the time domain overlap-add signal composition technique.

For the proposed HnH-NRSE, the harmonic factors and exponent coefficients are defined as $f_\text{HnH}=1.1$, $f'_\text{HnH}=0.7$, $c_\text{HnH}=0.2$ and $c'_\text{HnH}=0.1$.

\subsection{Proposed Composite Methods}
In this subsection, the composite methods IRMN and IRMO are introduced as a combination of the acoustic mask IRM and the speech enhancement methods NNESE and OMLSA.

$1)$ IRMN: The IRM \cite{Loizou_11} is defined for reverberant environments considering the SRR (Speech-to-Reverberant Ratio).
The main idea is to select TF regions on frame $l$ and band $j$ that have more energy related to the direct path signal and remove overlap-masking components due to reverberation inside a room.
Therefore, the IRM can be defined as
\begin{equation}
 \text{IRM}(l,j) =  \begin{cases} 
                1, \quad \text{SRR}(l,j) > \Theta_{Th} \\
                0, \quad \text{otherwise},
            \end{cases}
\end{equation}
where $\Theta_{Th}$ is a threshold in dB that defines the selection of TF regions.

The composite scheme proposed in this paper is based on the temporal speech enhancement method NNESE \cite{Tavares_16}.
In the NNESE procedure the noise standard deviation is estimated using the DATE estimator (\textit{d}-Dimensional Trimmed Estimator) \cite{Pastor_12} on a non-overlap frame-by-frame approach.  
To this end, a detection threshold $\xi(\rho_l)$ is defined as
\begin{equation}
    \xi(\rho_l) = \frac{1}{2}\rho_l + \frac{1}{\rho_l}\log \left( 1 + \sqrt{1 - \exp({-\rho_l^2})} \right)
    \label{eq:nnese1}
\end{equation}
where $\rho_l$ is the minimum SNR threshold for the $l$-th frame. 
For a Gaussian noise, $\rho_l=4$ and $\xi(\rho)=3.4742$.
 
The DATE estimator organize the $T$ samples of the $l$-th frame $y_l(t)$ by amplitude as $Y_1 \leq Y_2 \leq ... \leq Y_T$.
The Bienayme-Chebyshev-Markov inequality is then applied to calculate $t_{min}$ such that the samples lower than $Y_{t_{min}}$ are considered noise-only amplitudes.
\begin{equation}
    t_{min} = \frac{T}{2} - \frac{1}{\sqrt{4T(1-Q)}}T
    \label{eq:nnese2}
\end{equation}
where $Q$ is the confidence degree, which is assumed to be $95$\% for a Gaussian noise.

Then, it is verified the existence of a $b_l \in \{t_{min}, \dots, T\}$ that satisfies the relation
\begin{equation}
||Y_{t-1}|| \leq \frac{c \sum_{i=1}^{b_l} ||Y_i||}{t} \leq ||Y_{t+1}||,  
\end{equation}
where $c$ is adjustment factor that is dependent on the detection threshold:
\begin{equation}
 c=\frac{\xi(\rho) \Gamma(0.5)}{\sqrt{2}\Gamma(1.0)}. 
\end{equation}
Otherwise, it is considered to be $b_l = t_{min}$.
The noise standard deviation for that frame is then calculated as
\begin{equation}
	\hat{\sigma}_l = \frac{c \sum_{i=1}^{b_l} ||Y_i||}{b_l}.
	\label{eq:nnese}
\end{equation}

The amplitude is modified directly from the corrupted signal considering attenuation factors $\alpha$ and $\beta$ as 
\begin{equation}
    \Tilde{y}_l(t) = \begin{cases}
                        y_l(t) - \alpha \hat{\sigma}_l & , y_l(t) > Y(b_l)\\
                        \beta y_l(t) & , \text{otherwise}.
                    \end{cases}
    \label{eq:attnnese}
\end{equation}
The IRM is directly considered for the estimation of $\hat{\sigma_l}$.
After organizing the samples of a frame $l$, the standard deviation estimate is selected by the fraction threshold $\Phi_l$ of regions dominated by noisy-reverberant distortions in the local acoustic mask. 
For a total number of frequency bands $J$ adopted in the IRM, $\Phi_l$ is defined as
\begin{equation}
	\Phi_l = 1 - \frac{ \sum_{j=1}^{J}  \text{IRM}(l,j) }{J}, \quad \Phi_l \in [0,1].
\end{equation}
This way, the value $b_l$ in equation (\ref{eq:nnese}) is replaced by $b_l = \lfloor T . \Phi_l \rfloor$, where $\lfloor \bullet \rfloor$ is the integer operator.

After the amplitude removal of the distortion components, the acoustic mask is applied to the enhanced signal.
As the NNESE speech enhancement is temporal designed, the posterior application of IRM can deal with spectro-temporal regions that were not completely considered in a intrusive fashion, leading to an overall better speech intelligibility and quality improvement result.

$2)$ IRMO: The second comparative method is a variation of \cite{Lightburn_17} for environments with the reverberation effect.
The proposed adapted strategy adopts the acoustic mask IRM and OMLSA speech enhancement approaches and is  denominated as IRMO.

An important step of OMLSA technique is the association of the SSPs (Speech Presence Probabilities) in a given TF region.
This is performed in order to calibrate the gain for the enhancement procedure.
In this work, this step is replaced by a IRM indication of a TF speech region as follows.  

Consider $X(l,j)$ and $Y(l,j)$ the respective STFT coefficients of the direct and corrupted speech signals obtained via overlapping Hamming analysis windows.
Let the presence and absence of speech be considered under hypothesis $H_1(l,j)$ and $H_0(l,j)$, respectively.
Taking into account the assumption of statistically independent Gaussian random variables for the STFT coefficients, the desired gain $G(l,j)$ is defined such that 
\begin{equation}
    G(l,j)|Y(l,j)| = \exp\{E[\log|X(l,j)||Y(l,j)|]\},
\end{equation}
where $E[\cdot]$ is the expectation operator.
The OMLSA adopts the IMCRA \cite{cohen_2003} to estimate the noise power spectrum. 
After the estimation, this method reconstructs the enhanced speech signal by minimizing the mean-square error of the log-spectral amplitude.
Under the constraint of $G_{H_1}(l,j)$ assuming greater values than a threshold $G_{min}$ when speech is absent it is shown \cite{Cohen_01} that
\begin{equation}
    G(l,j) = {G_{H_1}(l,j)}^{p(l,j)} G_{min}^{q(l,j)},
\end{equation}
where $G_{H_1}(l,j)$ is the gain under the hypothesis that speech is present in the TF bin, $p(l,j)$ is the SPP and $q(l,j) = 1 - p(l,j)$ is the probability of absence of speech.

For the IRMO composite scheme, the values of $q(l,j)$  and $G_{min}$ are both determined by IRM$(l,j)$, as in
\begin{equation}
    q(l,j) = \begin{cases} 
                Q^1, \quad \text{IRM}(l,j) = 1\\
                Q^0, \quad \text{IRM}(l,j) = 0,
            \end{cases}
\end{equation}
and
\begin{equation}
    G_{min}(l,j) = \begin{cases} 
                G^1, \quad \text{IRM}(l,j) = 1\\
                G^0, \quad \text{IRM}(l,j) = 0,
            \end{cases}
\end{equation}
where $Q^1$, $Q^0$, $G^1$ and $G^0$ are defined in \cite{Lightburn_17}. 
The composition of these two methods merges the spectral cues present in the mask that are important to intelligibility with the quality gain imposed by the speech enhancement method.
This way, the approach can obtain simultaneous intelligibility and quality improvement under noisy-reverberant conditions.
Additionally, the reverberant mask threshold $\theta_{Th}$ is set to $-6$ dB for composite solutions IRMO and IRMN.


\section{Objective Evaluation of Speech Intelligibility and Quality}
This Section, briefly introduces of the objective measures applied in this work for speech intelligibility and speech quality prediction.


\subsection{Short-Time Intelligibility Objective Measure}
The short-time objective intelligibility measure (STOI) \cite{Taal_11} is a correlation-based method to compare the spectrum of the clean and the enhanced speech signals in the frequency domain. 
The correlation between temporal envelopes of the clean and noisy speech signals is defined as the intermediate intelligibility measure STOI$_{(j,l)}$ of each frequency band $j$ and each time frame $l$. 
The STOI is then computed as
\begin{equation}
 \text{STOI} = \frac{1}{15} \sum_{l=1}^L \sum_{j=1}^{15} \text{STOI}_{(j,l)} 
\end{equation}
where $L$ is the total number of speech frames.

%

\subsection{Short-Time Approximated Speech Intelligibility Index}
For approximated short-time versions of the SII \cite{ANSI_97}\cite{Hen_15}, time-frequency SNR $\xi(j, l)$ values are computed at each critical frequency band $j$ and time frame $l$, and than normalized to the $[0, 1]$ range.
The ASII\textsubscript{ST} measure assumes a smooth normalization for the time-frequency SNR, given by
\begin{equation}
\small
 d(j, l) = \frac{\xi(j, l)}{\xi(j, l)+1}.
  \label{Eq::ASII}
\end{equation}
The ASII\textsubscript{ST} is then computed as a weighted average of all values given in (\ref{Eq::ASII}):
\begin{equation}
\small
 \text{ASII\textsubscript{ST}} = \frac{1}{L} \sum_{l=1}^L \sum_{j=1}^{J} \gamma_j d(j,l) ,
\end{equation}
where $J$ is the total number of critical bands, and $\gamma_j$ are the critical-band-importance weights.
A relevant aspect of these short-time based measures is the capacity to account for the natural non-stationarity behavior of acoustic noises. 

%
%


\subsection{Perceptual Evaluation of Speech Quality}
The PESQ is an objective measure assess quality of narrow-banded speech and handset telephony \cite{Pesq_01}.
It is also largely adopted to estimate the quality of speech signals corrupted with noise due to its high correlation to hearing perceptual evaluation \cite{Loi_08}.

The PESQ score is calculated following four major steps of pre-processing, time alignment, loudness perception model transformation and disturbance computation.
Due to the non-stationary behavior of acoustic noises, in this work the symmetrical ($d_{s,l}$) and asymmetrical ($d_{a,l}$) distances are computed per frame between the analyzed and reference speech signals
\begin{equation}
    PESQ_l = 4.5 - 0.1 d_{s,l} - 0.0309 d_{a,l},
\end{equation}
such that the final PESQ score is average over all frames taking into account the variability of acoustic distortions.

\subsection{Speech-to-Reverberation Modulation Energy Ratio}
The non-intrusive SRMR quality metric \cite{Falk_14} estimates the human perceived reverberation effect on speech signals. 
This measure uses long-term temporal dynamics information to identify the presence of reverberation considering a double modulation per frame approach.
An average per-modulation band energy $\bar{\varepsilon_{j}}$ is obtained from Hilbert temporal envelopes. 
The SRMR is then computed as
\begin{equation}
\text{SRMR} = \frac{\sum_{j=1}^4 \bar{\varepsilon_{j}}}{\sum_{j=5}^{J^*} \bar{\varepsilon_{j}}}, 
\end{equation}
where $J^*$ is the upper summation bound that is dependent on the speech signal under evaluation.

\subsection{Perceptual Evaluation of Audio Quality}
The PEAQ \cite{PEAQ_1999} is a measurement of perceptual evaluation of coded audio signals based on several perceptual features called Model Output Variables (MOVs).
These features are typically computed estimated masking thresholds for the error signal or comparing the internal ear representation between processed and reference signals.
The measure is finally obtained using the output of a neural network.
In this work, the PEAQ f2-model variation is adopted with newly parameters and two MOVs: the Average Distorted Blocks (ADB) and the Average Modulation Difference \#1 (AvgModDiff1) as in \cite{PEAQ_2fmodel}.

\section{Experiments and Discussion} 
The speech intelligibility and quality results obtained with proposed solutions and baseline techniques ARA\textsubscript{NSD} \cite{Zuca_20} and TSDL \cite{Wang_2019} are presented in this Section.
Each testing scenario considered $100$ speech signals from the IEEE speech database \cite{Loizou_17}.
A total of $600$ speech signals is selected for the TSDL training step, with $100$ of these signals being used as validation set.
Acoustic noises Babble and Cafeteria were extracted from Freesound\footnote{Available at www.freesound.org.} database identified as ``Group Talking'' and ``Cafe Ambiance'', respectively.
Speech utterances have sampling rate of $16$ kHz and average time duration of $3$ seconds.

Several noisy-reverberant conditions are used to evaluate the proposed techniques in terms of speech intelligibility and quality.
In order to compose the noisy-reverberant environment, two real reverberation rooms (LASP2 and Stairway) and two non-stationary acoustic noises (Babble and Cafeteria) are considered in the experiments.
The LASP2 room is part of the LASP\_RIR\footnote{Available at lasp.ime.eb.br.} database with a reverberation time of $T_{60} = 0.79$ s. 
Room Stairway is selected from the AIR database \cite{Air_09} and presents reverberation times of $T_{60} = 1.1$ s.
The Babble and Cafeteria additive background noises are selected, respectively, from the RSG-10 \cite{Steen_88} and DEMAND \cite{Thie_2013} databases.
These noises are classified as non-stationary, with maximum INS of $40$ and $15$.


\begin{table}[t!]
\vspace{-.3cm}
  \caption{STOI results for LASP2 room and acoustic noises Babble and Cafeteria.}
  \centering
  \renewcommand{\tabcolsep}{0.9mm}
  \renewcommand*{\arraystretch}{1.2}
\resizebox{\columnwidth}{!}{
\begin{tabular}{c c c c c c c c} 	
\hline
Noise  & SNR & UNP & ARA\textsubscript{NSD} & TSDL & HnH-NRSE & IRMO & IRMN  \\ \hline 
\centering  \multirow{7}{*}{ \shortstack{Babble \\ INS $=40$} } 
& 2.3 &  64.0 & 64.2 & 64.6 & \bf67.1 & 85.1 & \bf87.0 \\ 
& 0.2 &  60.0 & 60.1 & 61.1 & \bf64.0 & 83.6 & \bf85.2 \\ 
& -1.9 & 56.0 & 56.8 & 58.4 & \bf61.0 & 81.5 & \bf83.1 \\ 
& -4.3 & 52.0 & 52.2 & 54.6 & \bf57.9 & 79.0 & \bf80.6 \\ 
& -7.3 & 48.0 & 48.6 & 50.1 & \bf54.8 & 75.6 & \bf77.3 \\ 
&-10.6 & 44.0 & 44.7 & 44.8 & \bf51.7 & 70.3 & \bf71.6 \\ 
& Average &  54.0 & 54.4 & 55.6 & \bf59.4 & 79.2 & \bf80.8 \\ \hline 
\centering  \multirow{7}{*}{ \shortstack{Cafeteria \\ INS $=15$}  } 
&1.3 &  64.0 & 64.1 & 65.8 & \bf66.4 &  85.3 & \bf87.4 \\ 
&-0.8 & 60.0 & 60.2 & \bf63.7 & 63.4 &  83.7 & \bf85.8 \\ 
&-2.9 & 56.0 & 56.2 & \bf61.3 & 60.3 &  81.6 & \bf83.8 \\ 
&-5.4 & 52.0 & 52.7 & \bf58.7 & 57.2 &  79.1 & \bf81.4 \\ 
&-8.3 & 48.0 & 48.3 & 53.7 & \bf54.1 &  75.6 & \bf77.8 \\ 
&-12.2 &44.0 & 41.1 & 46.8 & \bf51.0 &  70.4 & \bf73.6  \\ 
&Average &  54.0 & 54.3 & 58.3 & \bf58.7 &  79.3 & \bf81.6 \\ \hline 
\multicolumn{2}{r}{ Overall Average} & 54.0 & 54.4 & 57.0 & \bf 59.1 & 79.2 & \bf 81.4
\\ \hline
\end{tabular}
}
\label{Tab::STOI_Lasp2} 
\end{table}  
\begin{table}[t!]
\vspace{-.3cm}
  \caption{STOI results for Stairway room and acoustic noises Babble and Cafeteria.}
  \centering
  \renewcommand{\tabcolsep}{0.9mm}
  \renewcommand*{\arraystretch}{1.2}
\resizebox{\columnwidth}{!}{
\begin{tabular}{c c c c c c c c} 	
\hline
Noise  & SNR & UNP & ARA\textsubscript{NSD} & TSDL & HnH-NRSE & IRMO & IRMN  \\ \hline 
\centering  \multirow{7}{*}{ \shortstack{Babble \\ INS $=40$} } 
&  4.2 & 64.0 & 64.3 & 63.5 & \bf68.6 & 86.9 & \bf89.0 \\ 
&  2.1 & 60.0 & 60.4 & 61.2 & \bf65.4 & 85.6 & \bf87.4 \\ 
&  0.1 & 56.0 & 56.0 & 58.6 & \bf62.2 & 83.9 & \bf85.5 \\ 
& -2.2 & 52.0 & 52.7 & 55.4 & \bf59.0 & 81.7 & \bf83.1 \\ 
& -4.8 & 48.0 & 48.3 & 51.3 & \bf55.7 & 78.9 & \bf80.0 \\ 
& -8.6 & 44.0 & 44.8 & 45.9 & \bf52.3 & 74.1 & \bf75.4 \\ 
& Average & 54.0 & 54.4 & 56.0 & \bf60.5 & 81.8 & \bf83.4  \\ \hline 
\centering  \multirow{7}{*}{ \shortstack{Cafeteria \\ INS $=15$}  } 
&  2.5 & 64.0 & 64.2 & 66.2 & \bf68.3 & 87.1 & \bf89.0  \\ 
&  0.6 & 60.0 & 60.2 & 64.2 & \bf65.2 & 85.6 & \bf87.4  \\ 
& -1.5 & 56.0 & 56.9 & 61.5 & \bf62.1 & 83.8 & \bf85.5  \\ 
& -3.8 & 52.0 & 52.6 & 57.9 & \bf58.9 & 81.6 & \bf83.1  \\ 
& -6.5 & 48.0 & 48.4 & 53.2 & \bf55.6 & 78.4 & \bf79.8  \\ 
&-12.8 & 44.0 & 44.1 & 47.2 & \bf52.3 & 73.2 & \bf75.6  \\ 
& Avg. & 54.0 & 54.4 & 58.4 & \bf60.4 & 81.6 & \bf83.4 \\ \hline 
\multicolumn{2}{r}{ Overall Average} & 54.0 & 54.4 & 57.2 & \bf 60.5 & 81.7 & \bf 83.4
\\ \hline
\end{tabular}
}
\label{Tab::STOI_Stairway} 
\end{table}

\begin{figure}[t!]
    \centering
    \includegraphics[width=8.2cm]{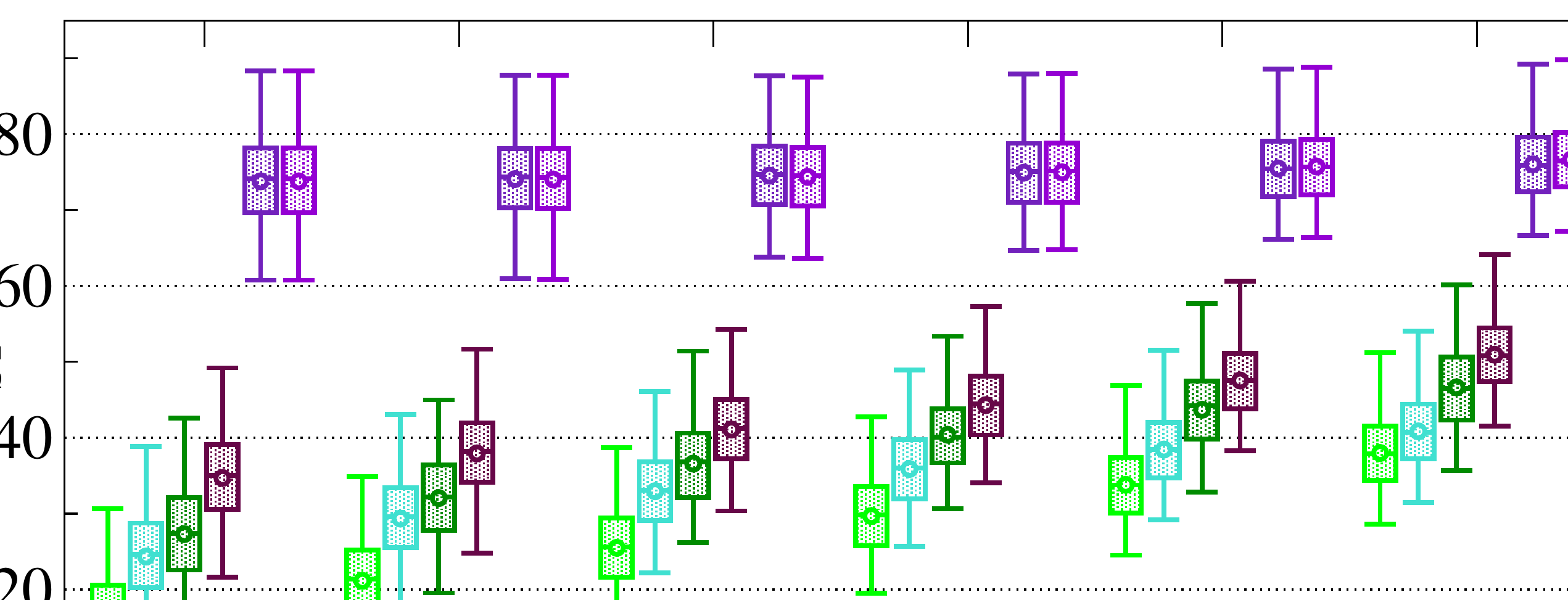} \\
    \vspace{1.3cm}
    {\centering \small \hspace{0.5cm} (a)} \\
    \vspace{0.5cm}
    \includegraphics[width=8.2cm]{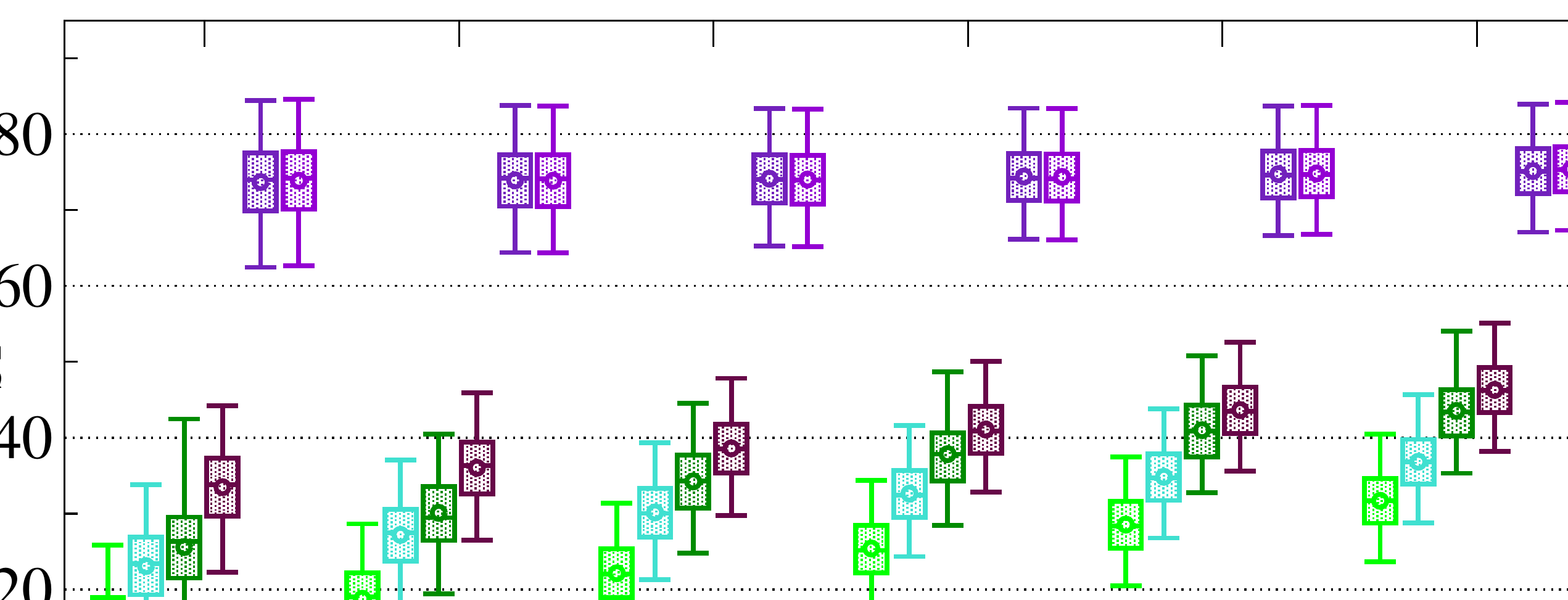} \\
    \vspace{1.3cm}
    {\centering \small \hspace{0.5cm} (b)} \\
    \caption{Boxplot of ASII\textsubscript{ST} objective values for (a) LASP2 room and (b) Stairway room, considering acoustic noises Babble and Cafeteria}
    \label{Fig::ASII_Lasp}
\end{figure} 


Speech signals are corrupted considering SNRs values ranging from $-12.8$ dB up to $4.2$ dB, where the SNRs are measured between the reverberant speech signal and the background noise.
For each reverberant signal, SNRs are selected in order to obtain STOI \cite{Taal_11} objective intelligibility scores of \{$0.44$, $0.48$, $0.52$, $0.56$, $0.60$ and $0.64$\}.
The limit values of $0.44$ and $0.64$ are defined here as thresholds of poor and good intelligibility for the unprocessed (UNP) noisy-reverberant speech.
Results are organized considering the increasing value of $T_{60}$, i.e., for each measure rooms LASP2 ($T_{60} = 0.79$ s) and Stairway ($T_{60} = 1.1$ s) are presented consecutively with acoustic noises Babble and Cafeteria within each case.

A listening test \cite{Ghimire_12} with $10$ native Brazilian volunteers ($5$ male and $5$ female) was conducted adopting a closed scenario of phonetic balanced words\footnotemark[2].
Their ages ranged from $20$ to $24$ years with an average of $22$.
A simulated room with $7.0$ x $5.2$ x $3.0$ m$^3$ and $T_{60}=1.0$ s was generated by the image source method (ISM) \cite{Allen_79}.
The SSN acoustic noise was adopted with SNRs of $-3$ dB, $0$ dB and $3$ dB.
$10$ words were considered for each of test conditions, i.e., $3$ SNR levels for $5$ methods plus the unprocessed case.  
Participants were introduced to the task in a training session with $4$ words.
The material was diotically presented using a pair of Sennheiser HD200 headphones. 
Listeners heard each word once in an arbitrary presentation order and selected one among five words in order to compute the Word Recognition Rate (WRR).

\subsection{Objective Speech Intelligibility Evaluation}
The proposed and competing methods are first evaluated in terms of STOI intelligibility scores.
Tables \ref{Tab::STOI_Lasp2} and \ref{Tab::STOI_Stairway} show the STOI values for rooms LASP2 and Stairway, respectively.
Note that, the HnH-NRSE attains the highest STOI average results for most SNR conditions and noises when considering the ARA\textsubscript{NSD} and TSDL methods.
This is particularly true for the Stairway room with higher $T_{60}$, where the proposed solution surpasses the competing methods in all cases.
The HnH-NRSE acquires an average STOI value of $59.1$ and $60.5$ for rooms LASP2 and Stairway.
Equivalent intelligibility scores of $57.0$ and $57.2$ are observed for the TSDL whereas $54.5$ and $54.4$ are achieved by the ARA\textsubscript{NSD}.
The lower intelligibility acquired by the ARA\textsubscript{NSD} for the STOI measure can be explained by its processing approach that absorbs frames with dominant masking components without considerations on the spectral envelope correlation which are relevant for the STOI intelligibility measures.
The highest overall average intelligibility results are achieved by composed solutions IRMO and IRMN.
For theses reference approaches, the proposed IRMN is able to acquire the highest intelligibility scores for all scenarios.
The IRMN attained, an average STOI of $81.4$ and $83.4$ for rooms LASP2 and Stairway, respectively.
This indicates intelligibility gains of $27.4$ and $29.4$ for the IRMN solution against $25.2$ and $27.1$ for the IRMO technique.

Figure \ref{Fig::ASII_Lasp} depicts the box-plot of ASII\textsubscript{ST} intelligibility scores for rooms LASP2 and Stairway considering both acoustic noises Babble and Cafeteria.
Note that for this objective measure, the HnH-NRSE surpasses methods ARA\textsubscript{NSD} and TSDL in all conditions, achieving the highest average ASII\textsubscript{ST} results.
In light of these solutions, the variance of ASII\textsubscript{ST} values are similar for all STOI scenarios.
For the LASP2 room, the proposed HnH-NRSE overall average is $42.8$ in comparison with $37.8$ for the TSDL and $29.2$ for the ARA\textsubscript{NSD}.
Equivalent scores for Stairway room are $39.9$, $35.4$ and $26.0$, respectively.
This implies an overall average intelligibility gain of $15.9$ for HnH-NRSE, $11.1$ for TSDL and $6.8$ for ARA\textsubscript{NSD}, which corroborates the capacity of the proposed method to increase speech intelligibility under noisy-reverberant conditions.
Furthermore, this indicates that the proposed approach is suited for real acoustic effects with different non-stationary behavior.
For the composed techniques, the IRMN attained overall averages of $74.9$ and $74.4$ for rooms LASP2 and Stairway in comparison to $74.8$ and $74.4$ for the IRMO, respectively.
The similarity of these results can be partially explained by the direct adoption of local time-frequency SNR for the ASII\textsubscript{ST} computation and the usage of the IRM acoustic mask for both composite measures, as the IRM imposes on the enhanced speech signal the spectral cues present in the mask that are important to intelligibility.



\vspace{-.2cm}
\subsection{Perceptual Intelligibility Evaluation}

\begin{figure}[t!]
    \centering
    \vspace{.2cm}
    \includegraphics[width=\linewidth]{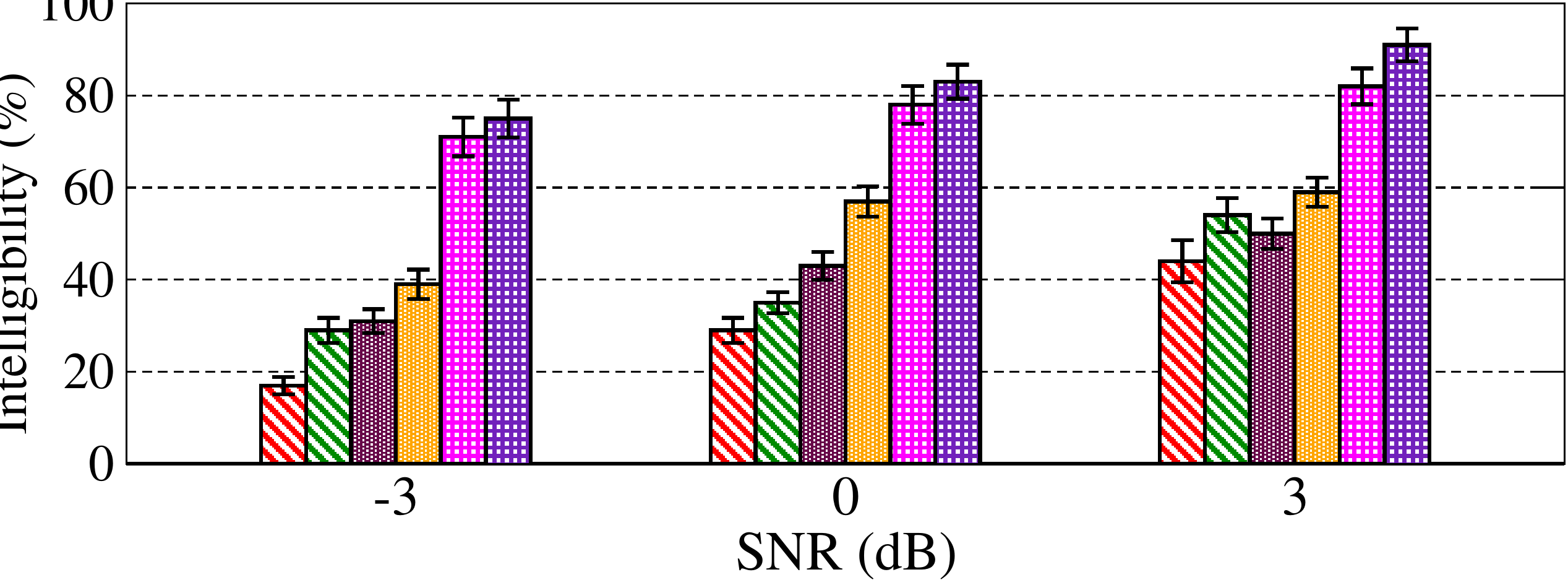}
    \vspace{-.7cm}
    \caption{Intelligibility results and standard deviation for perceptual evaluation.}
    \label{Fig::Percept_Eval}
\end{figure}

A perceptual evaluation is also performed in this work.
The average intelligibility scores and standard deviations values for each method are presented in Fig. \ref{Fig::Percept_Eval}.
The HnH-NRSE improves the intelligibility under all conditions over competing techniques. 
The proposed method achieves intelligibility improvements of $22$ p.p., $28$ p.p. and $15$ p.p. against $14$ p.p., $14$ p.p. and $6$ p.p. for the TSDL and $12$ p.p., $6$ p.p. and $10$ p.p. for ARA\textsubscript{NSD} under SNR values of $-3$ dB, $0$ dB and $3$ dB, respectively. 
Furthermore, composed solutions attained overall average intelligibility improvements of $47$ for IRMO and $53$ for IRMN.
This result corroborates with previous objective findings on the proposed strategies capacity to provide intelligibility improvement.

\begin{table}[t!]
\vspace{-.3cm}
 \caption{PESQ results for LASP2 room and acoustic noises Babble and Cafeteria.}
 \centering
 \renewcommand{\tabcolsep}{0.9mm}
 \renewcommand*{\arraystretch}{1.2}
\resizebox{\columnwidth}{!}{
\begin{tabular}{c c c c c c c c} 	
\hline
Noise & SNR & UNP & ARA\textsubscript{NSD} & TSDL & HnH-NRSE & IRMO & IRMN \\ \hline 
\centering \multirow{7}{*}{ \shortstack{Babble \\ INS $=40$} } 
& 2.3  & 1.93 & 1.79 & 1.95 & \bf2.01 & 3.02 & \bf3.02 \\ 
& 0.2  & 1.85 & 1.65 & 1.87 & \bf1.90 & \bf2.90 & 2.87 \\ 
& -1.9 & 1.72 & 1.69 & \bf1.82 & 1.78 & \bf2.74 & 2.73 \\ 
& -4.3 & 1.64 & 1.61 & \bf1.70 & 1.68 & 2.58 & \bf2.58 \\ 
& -7.3 & 1.50 & 1.60 & \bf1.63 & 1.57 & 2.38 & \bf2.41 \\ 
&-10.6 & 1.44 & 1.58 & 1.41 & \bf1.49 & 2.16 & \bf2.18 \\ 
& Average & 1.68 & 1.66 & 1.72 & \bf1.74 & 2.63 & \bf2.63 \\ \hline 
\centering \multirow{7}{*}{ \shortstack{Cafeteria \\ INS $=15$} } 
&1.3 & 1.95 & 1.97 & 1.98 & \bf2.10 & \bf3.12 & 3.11 \\ 
&-0.8 & 1.90 & 1.98 & 1.92 & \bf1.99 & \bf3.00 & 2.98 \\ 
&-2.9 & 1.84 & 1.86 & 1.87 & \bf1.88 & \bf2.86 & 2.84 \\ 
&-5.4 & 1.76 & 1.77 & \bf1.82 & 1.77 & 2.69 & \bf2.69 \\ 
&-8.3 & 1.56 & 1.74 & \bf1.78 & 1.67 & 2.51 & \bf2.52 \\ 
&-12.2 & 1.54 & 1.63 & \bf1.62 & 1.58 & 2.29 & \bf2.31 \\ 
&Average & 1.76 & 1.82 & 1.83 & \bf1.83 & 2.74 & \bf2.74 \\ \hline 
\multicolumn{2}{r}{ Overall Average} & 1.72 & 1.74 & 1.77 & \bf 1.78 & 2.68 & \bf 2.68
\\ \hline
\end{tabular}
}
\label{Tab::PESQ_Lasp2} 
\end{table} 
\begin{table}[t!]
\vspace{-.3cm}
  \caption{PESQ results for Stairway room and acoustic noises Babble and Cafeteria.}
  \centering
  \renewcommand{\tabcolsep}{0.9mm}
  \renewcommand*{\arraystretch}{1.2}
\resizebox{\columnwidth}{!}{
\begin{tabular}{c c c c c c c c} 	
\hline
Noise  & SNR & UNP & ARA\textsubscript{NSD} & TSDL & HnH-NRSE & IRMO & IRMN  \\ \hline 
\centering  \multirow{7}{*}{ \shortstack{Babble \\ INS $=40$} } 
&  4.2 	  & 2.00 & 1.97 & 2.03 & \bf2.26 & 3.12 & \bf3.12 \\ 
&  2.1 	  & 1.93 & 1.94 & 1.99 & \bf2.13 & \bf3.03 & 3.02 \\ 
&  0.1 	  & 1.90 & 1.88 & 1.95 & \bf2.01 & \bf2.92 & 2.90 \\ 
& -2.2 	  & 1.81 & 1.85 & \bf1.93 & 1.89 & \bf2.79 & 2.78 \\ 
& -4.8 	  & 1.71 & 1.82 & \bf1.84 & 1.78 & 2.63 & \bf2.65 \\ 
& -8.6 	  & 1.66 & 1.78 & \bf1.74 & 1.68 & 2.48 & \bf2.48 \\ 
& Average & 1.83 & 1.87 & 1.91 & \bf1.96 & 2.83 & \bf2.83 \\ \hline 
\centering  \multirow{7}{*}{ \shortstack{Cafeteria \\ INS $=15$}  } 
&  2.5    & 2.10 & 2.12 & 2.26 & \bf2.39 & 3.11 & \bf3.11  \\ 
&  0.6    & 2.03 & 2.10 & 2.22 & \bf2.27 & \bf3.03 & 3.02  \\ 
& -1.5    & 2.00 & 2.01 & 2.16 & \bf2.16 & \bf2.93 & 2.91  \\ 
& -3.8    & 1.94 & 1.93 & \bf2.11 & 2.04 & \bf2.80 & 2.79  \\ 
& -6.5    & 1.91 & 1.90 & \bf2.06 & 2.01 & 2.63 & \bf2.66  \\ 
&-12.8    & 1.78 & 1.88 & \bf2.01 & 1.96 & 2.44 & \bf2.55  \\ 
& Average & 1.96 & 1.99 & 2.14 & \bf2.14 & 2.82 & \bf2.84  \\ \hline 
\multicolumn{2}{r}{ Overall Average} & 1.89 & 1.93 & 2.02 & \bf 2.05 & 2.82 & \bf 2.84
\\ \hline
\end{tabular}
}
\label{Tab::PESQ_Stairway} 
\end{table}

\subsection{Speech Quality Measures}

The PESQ quality results are presented in Tables \ref{Tab::PESQ_Lasp2} and \ref{Tab::PESQ_Stairway}.
In this case, the UNP speech signal attained an overall average quality of $1.72$ and $1.89$ for rooms 
LASP2 and Stairway. 
The highest scores are obtained by HnH-NRSE corresponding values of $1.78$ and $2.05$, followed by TSDL with $1.77$ and $2.02$ and ARA\textsubscript{NSD} with $1.74$ and $1.93$.
This indicates that the proposed method outperforms competing non-composite strategies for the majority of noisy-reverberant scenarios.
This can be particularly noted for higher SNR values, where the HnH-NRSE accomplishes the highest PESQ gains.
For the composite techniques, the IRMN and IRMO attained the same overall average $2.68$ for room LASP2.
Considering the most reverberant room Stairway, the IRMN achieves a better PESQ score of $2.83$ against $2.82$ for the IRMO.

\begin{figure*}[t!]
    \centering
    \includegraphics[width=8.4cm]{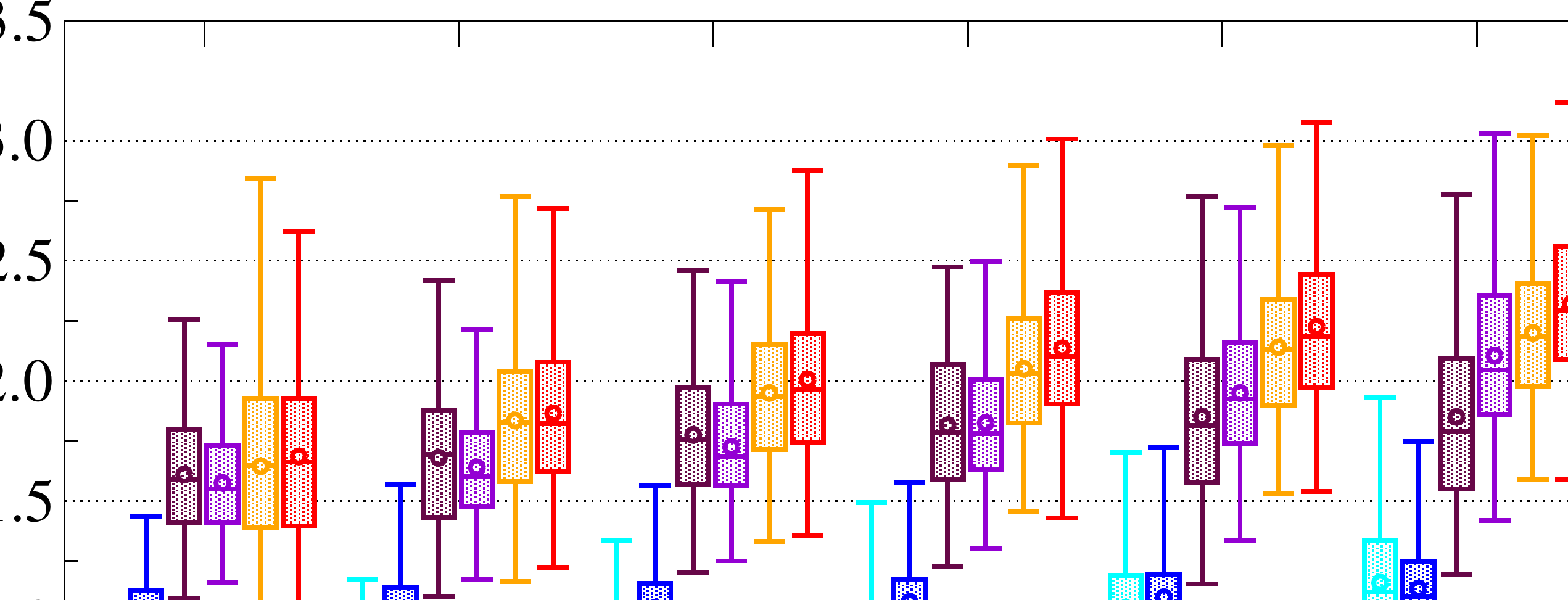} \hspace{0.6cm}
    \includegraphics[width=8.4cm]{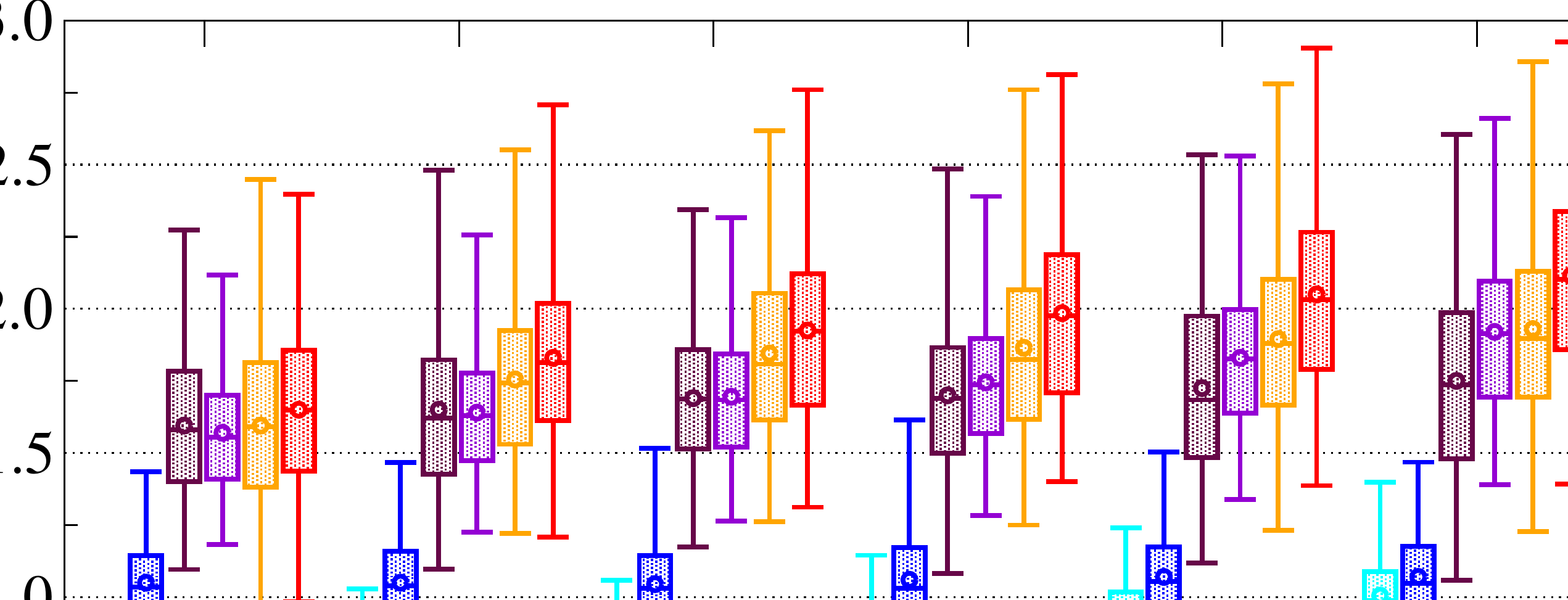} \\
    \vspace{1.3cm}
    {\small (a) \hspace{8.7cm} (b)} \\
    \vspace{-0.1cm}
    \caption{Boxplot of SRMR objective values for (a) LASP2 room and (b) Stairway room, considering acoustic noises Babble and Cafeteria}
    \label{Fig::SRMR}
\end{figure*}

In this work, proposed and competing methods are also evaluated in terms of SRMR quality scores.
Figure \ref{Fig::SRMR} depicts the SRMR box-plot values for the noisy-reverberant scenarios. 
Note that, the HnH-NRSE attains the highest SRMR results for most STOI conditions when considering the ARA\textsubscript{NSD} and TSDL methods.
This can be noted specially for higher STOI results.
The HnH-NRSE attains an average quality of $1.80$ for LASP2 room against $1.76$ and $1.05$ for TSDL and ARA\textsubscript{NSD}.
The same behavior can be noted for the Stairway room, with equivalent quality average scores of $1.73$, $1.68$ and $1.06$.
Note that the proposed HnH-NRSE solution attains SRMR quality scores that are close to composite approaches IRMO and IRMN.
This is particularly true for higher STOI scores. 
For the condition of STOI $=0.64$, the HnH-NRSE attains SRMR of $1.92$ which is the same as the composite method IRMO, which indicates that the proposed approach is able to improve quality of speech signals under adverse noisy-reverberant conditions.
Regarding composed methods, the IRMN attained the highest SRMR scores in all noisy-reverberant scenarios, with the  average scores of $2.04$ and $1.93$ for rooms LASP2 and Stairway, respectively.
The competing IRMO acquired $1.97$ and $1.81$ average SRMR values for the same acoustic environments.

\begin{figure}[t!]
    \centering
    \vspace{-0.2cm}
    \includegraphics[height=0.65cm]{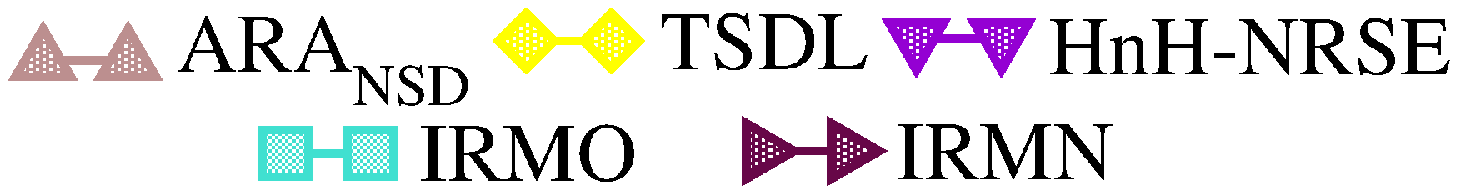} \\ \vspace{0.2cm}
    \includegraphics[width=4.3cm]{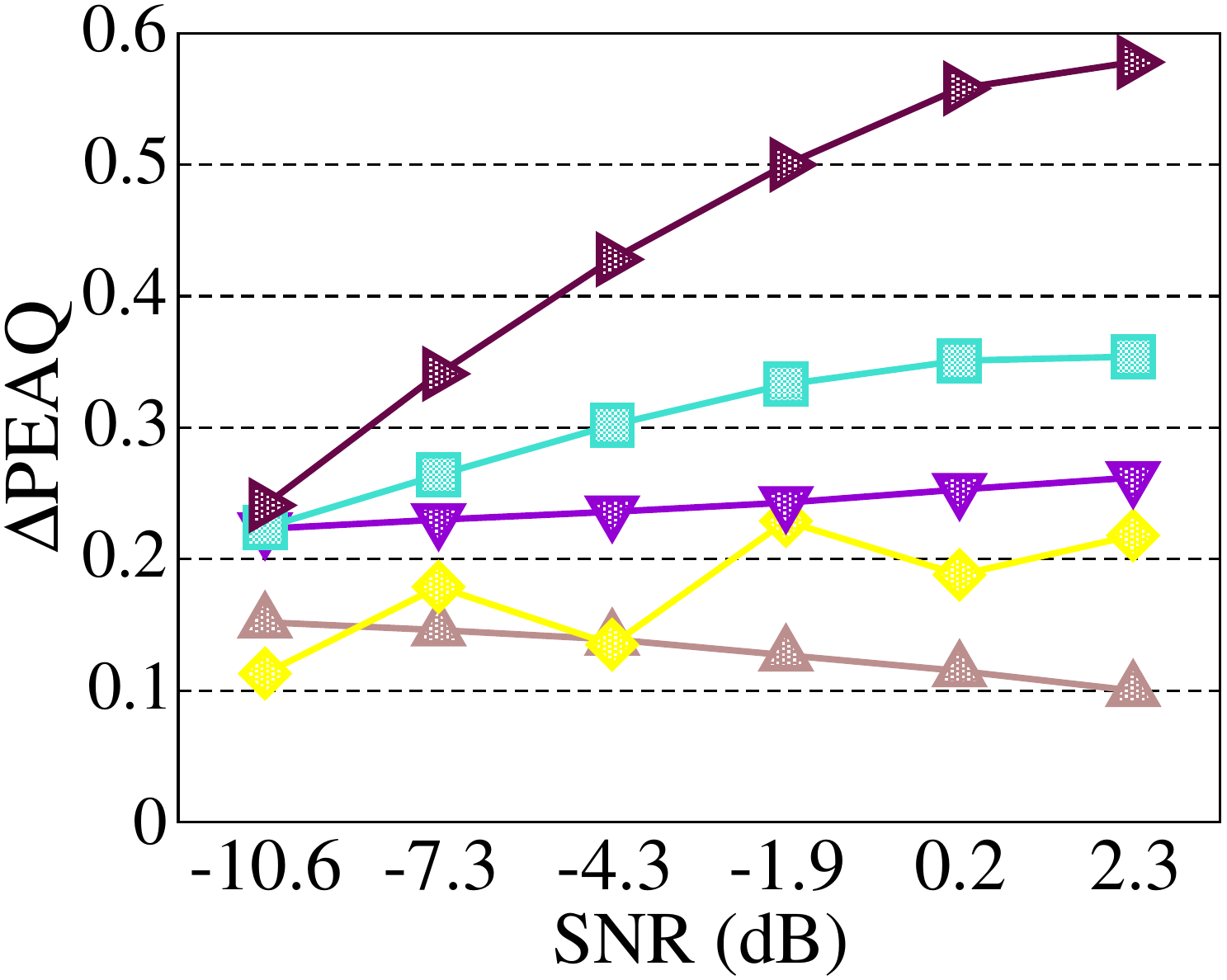}
    \includegraphics[width=4.3cm]{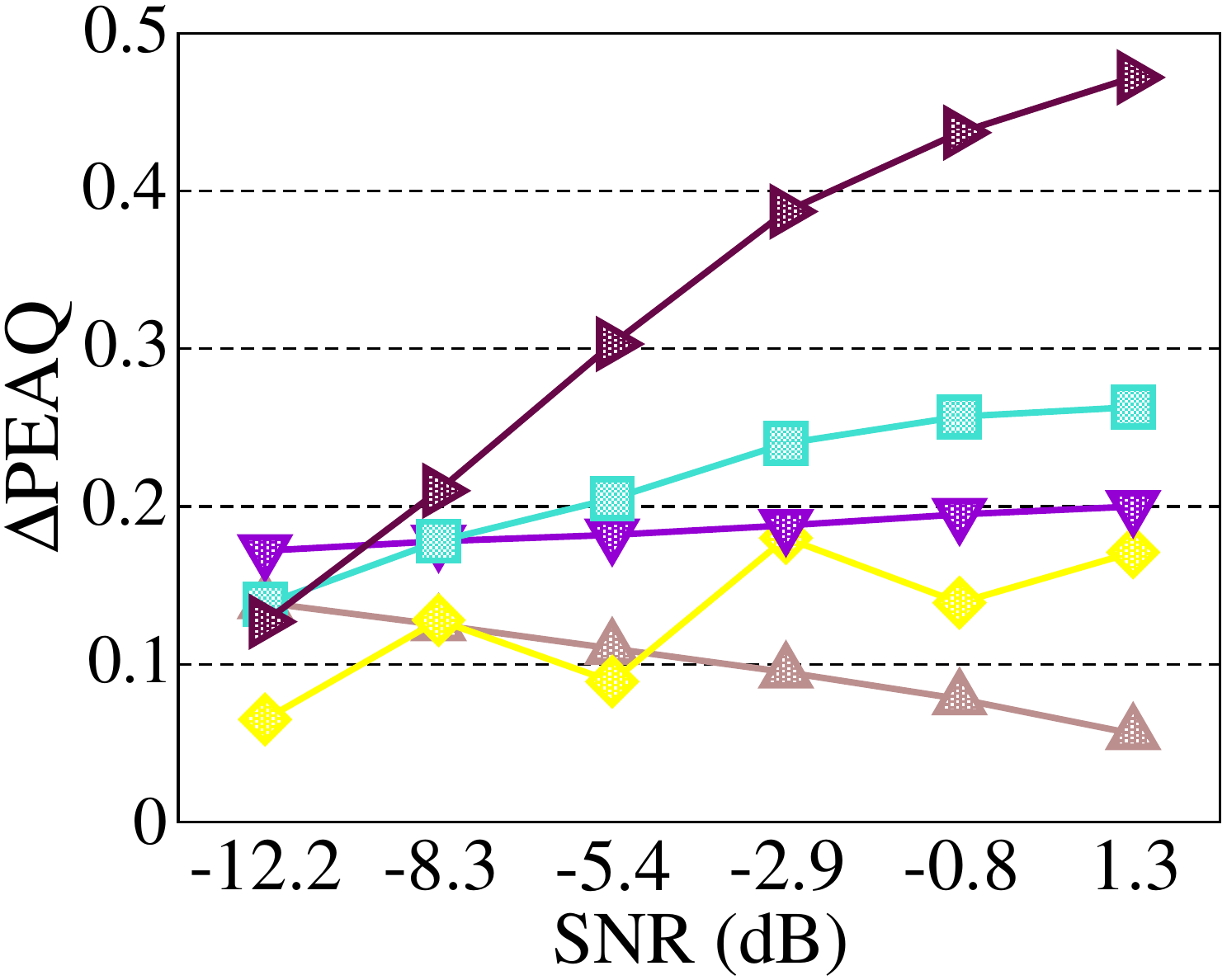} \\
    {\small  \vspace{-0.15cm} (a) \hspace{3.5cm} (b)} \\
    \caption{PEAQ quality improvement for LASP2 room with (a) Babble and (b) Cafeteria acoustic noises.}
    \vspace{0.1cm}
    \label{Fig::PEAQ_LASP}
\end{figure}

\begin{figure}[t!]
    \centering
    \vspace{-0.2cm}
    \includegraphics[height=0.65cm]{IMGs/Label_PEAQ_Cap4_sel_v2.eps} \\ \vspace{0.2cm}
    \includegraphics[width=4.3cm]{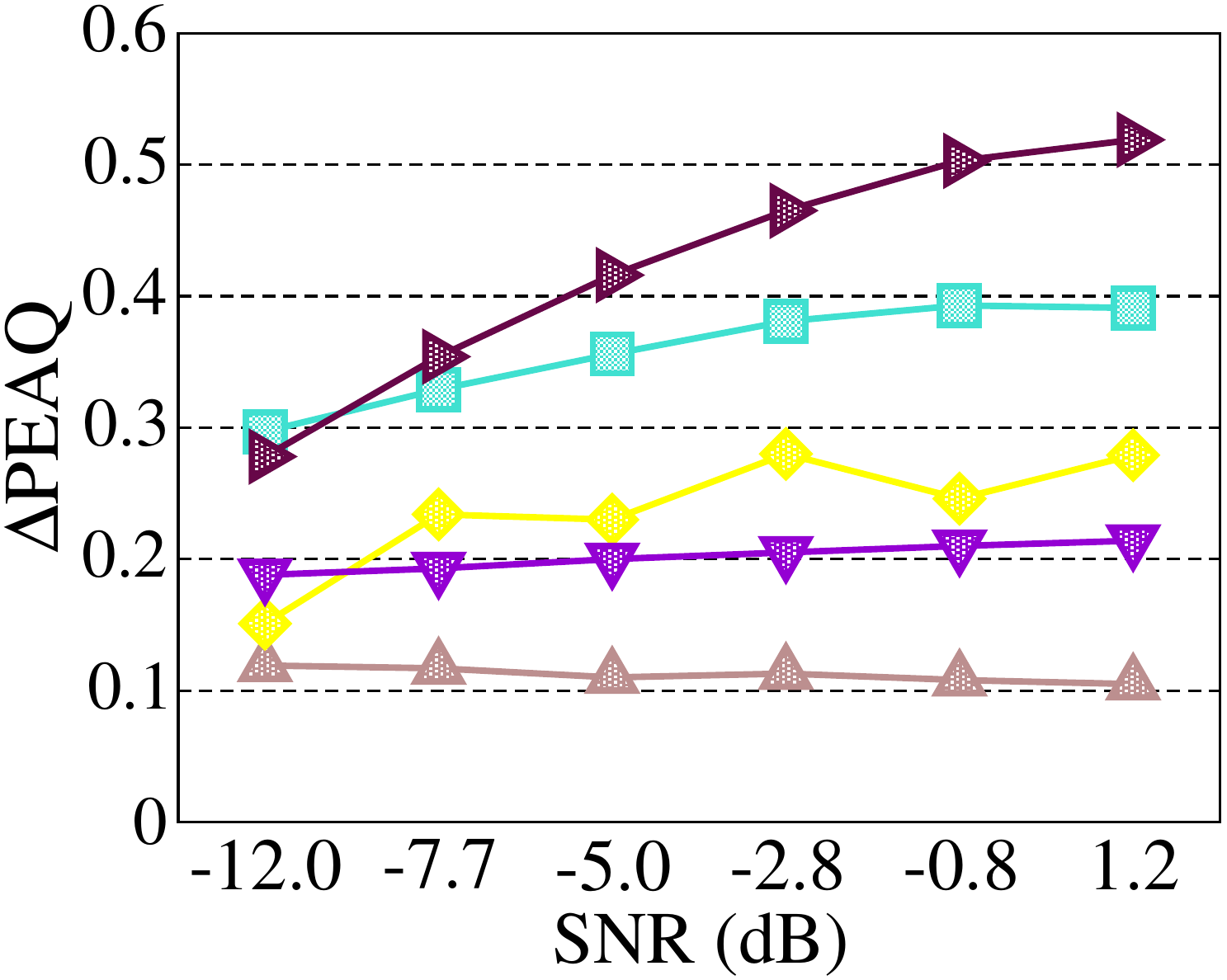}
    \includegraphics[width=4.3cm]{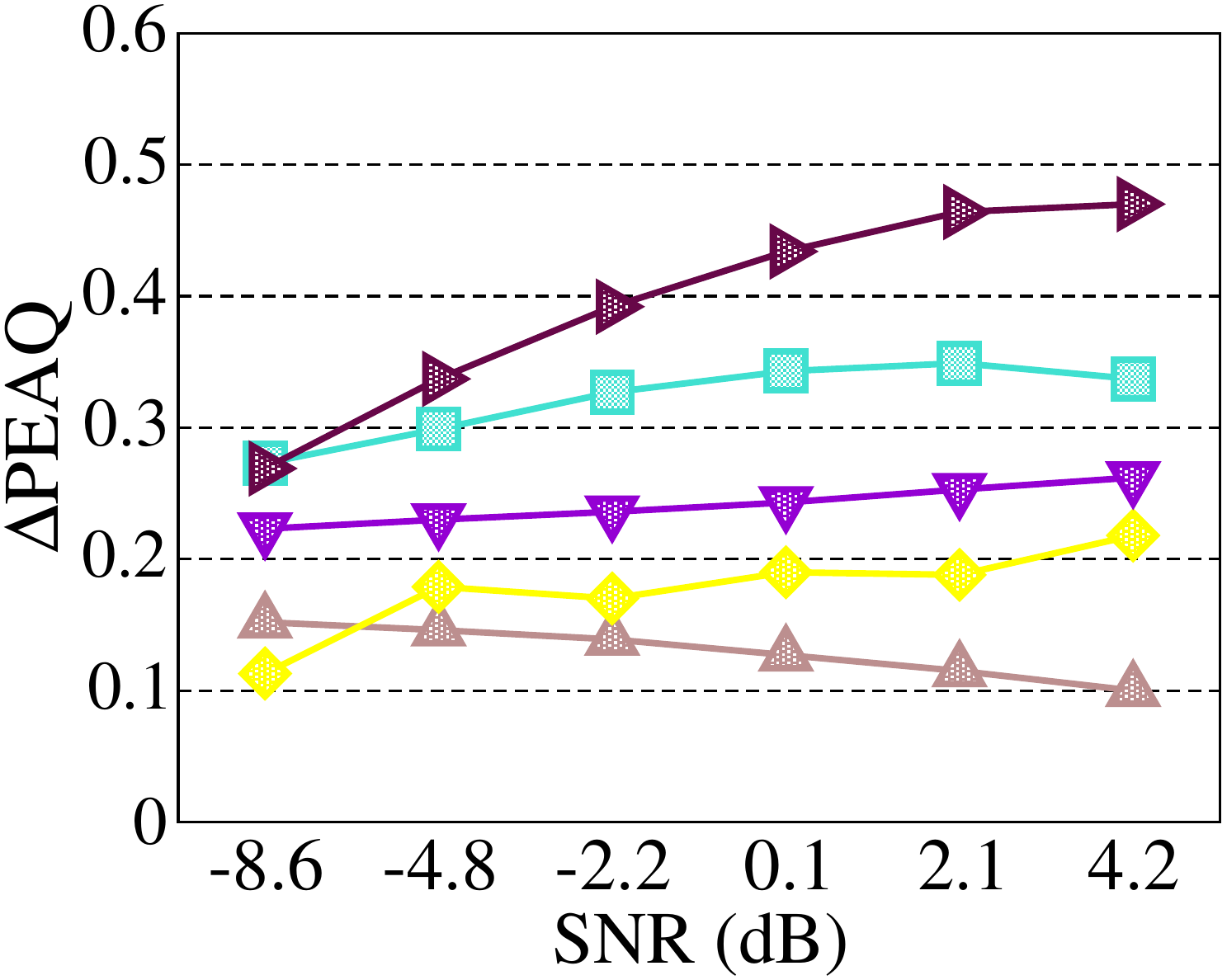} \\
    {\small  \vspace{-0.15cm} (a) \hspace{3.5cm} (b)} \\
    \caption{PEAQ quality improvement for Stairway room with (a) Babble and (b) Cafeteria acoustic noises.}
    \vspace{0.1cm}
    \label{Fig::PEAQ_Stairway}
\end{figure}

For the PEAQ objective measure, the quality improvement attained by each method is depicted in Figures \ref{Fig::PEAQ_LASP} and \ref{Fig::PEAQ_Stairway} for rooms LASP2 and Stairway, respectively.
The HnH-NRSE outperforms competing methods in most scenarios, leading to an overall average gain of  $0.21$ for LASP2 and $0.26$ for Stairway.
Considering the most non-stationary Babble noise and the greater $T_{60}$ of $1.1$ s for the Stairway room, the proposed method acquired the highest $\Delta$PEAQ gain in all SNR conditions, with an average $\Delta$PEAQ of $0.24$.
This result corroborates the capacity of the proposed method to deal with real acoustic effects of different non-stationary  natures, being able to provide both quality and intelligibility under diverse noisy-reverberant environments.
As means of comparison, the TSDL and ARA\textsubscript{NSD} strategies achieved average $\Delta$PEAQ values of $0.18$ and $0.13$ for the same condition.
Note that quality gains attained by HnH-NRSE are in most cases the closest to the composite methods IRMO and IRMN scores.
The IRMN obtains the highest quality improvements for most environments regarding composite methods. 
This reinforces that the combined usage of IRM and NNESE on noisy-reverberant scenarios simultaneously provide the best quality and intelligibility gains.

\section{Conclusion}
In this work, the HnH-NRSE method was proposed as a single time domain solution for simultaneous speech intelligibility and quality improvement under noisy-reverberant conditions.
To this end, the technique performs an harmonic decomposition and an objective assessment of the non-stationarity variations of speech signal in order to define an adaptive gain to deal with masking components.
Two other composite solutions  IRMO and IRMN were presented as combined methods for simultaneous improvement on noisy and reverberant scenarios, combining the IRM mask with OMLSA and NNESE speech enhancement solutions.
The HnH-NRSE attained an overall average intelligibility improvement of $5.8$ p.p. for the STOI measure and $15.9$ p.p. for the ASII\textsubscript{ST} measure.
The highest STOI improvement was achieved by the composed solution IRMN with $28.4$ p.p. intelligibility gain.
Objective quality results demonstrated that the HnH-NRSE is also able to increase speech quality, with the highest PESQ increment of $0.26$ for the Stairway room with Babble noise at SNR of $4.2$ dB. 
The proposed HnH-NRSE also obtained average SRMR quality scores of $1.92$ for the Stairway room, which is similar to the reference composed approache IRMO, whereas the IRMN attained $2.11$ in this scenario.
A perceptual intelligibility listening test was further considered and corroborated the objective results performed in this work.  

\bibliographystyle{IEEEtran}
\bibliography{BIBall}

\end{document}